\documentclass[aps,prd,twocolumn,showpacs,superscriptaddress,preprintnumbers]{revtex4-1}
\usepackage[colorlinks=true, pdfstartview=FitV, linkcolor=red,citecolor=blue, urlcolor=blue,
bookmarks=true, bookmarksnumbered=true, breaklinks]{hyperref}
\usepackage[dvipdfmx]{graphicx}
\usepackage{amsmath,amssymb,tabularx,ulem,mathrsfs,bm}
\usepackage{braket}
\begin{document}

\title{Heavy meson spectroscopy under strong magnetic field}

\author{Tetsuya Yoshida}
\email{t.yoshida@th.phys.titech.ac.jp}
\affiliation{Department of Physics, Tokyo Institute of Technology, Meguro, Tokyo, 152-8551, Japan}
\affiliation{Theoretical Research Division, Nishina Center, RIKEN, Wako, Saitama, 351-0198, Japan}

\author{Kei Suzuki}
\email{k.suzuki.2010@th.phys.titech.ac.jp}
\affiliation{Department of Physics and Institute of Physics and Applied Physics, Yonsei University, Seoul 03722, Korea}


\begin{abstract}
Spectra of the neutral heavy mesons, $\eta_c(1S,2S)$, $J/\psi$, $\psi(2S)$, $\eta_b(1S,2S,3S)$, $\Upsilon(1S,2S,3S)$, $D$, $D^\ast$, $B$, $B^\ast$, $B_s$ and $B_s^\ast$, in a homogeneous magnetic field are analyzed by using a potential model with constituent quarks. To obtain anisotropic wave functions and the corresponding eigenvalues, the cylindrical Gaussian expansion method is applied, where the wave functions for transverse and longitudinal directions in the cylindrical coordinate are expanded by the Gaussian bases separately. Energy level structures in the wide range of magnetic fields are obtained and the deformation of the wave functions is shown, which reflects effects of the spin mixing, the Zeeman splitting and quark Landau levels. The contribution from the magnetic catalysis in heavy-light mesons is discussed as a change of the light constituent quark mass.
\end{abstract}
\pacs{25.75.-q, 14.40.Pq, 12.39.-x}
\maketitle

\section{Introduction} \label{Sec_Introduction}
Modification of hadron properties in extreme environments such as finite temperature and density is one of the important problems in hadron physics, which is useful to clarify the relation between hadrons and the vacuum.
Similarly, an external magnetic field comparable to the typical scale of quantum chromodynamics (QCD) may drastically change hadron properties.
Such an intense field can be induced in noncentral heavy-ion collisions between two charged nuclei, where the strength is estimated to reach $|eB| \sim m_\pi^2 \sim 0.02 \, \mathrm{GeV^2}$ at the Relativistic Heavy Ion Collider (RHIC) and $|eB| \sim 15 m_\pi^2 \sim 0.3 \, \mathrm{GeV^2}$ at the Large Hadron Collider (LHC) \cite{Kharzeev:2007jp,Skokov:2009qp,Voronyuk:2011jd,Ou:2011fm,Bzdak:2011yy,Deng:2012pc,Bloczynski:2012en,Bloczynski:2013mca,Deng:2014uja,Huang:2015oca}.
Moreover, lattice QCD simulations are powerful tools to numerically confirm hadron properties in strong magnetic fields.

For light mesons (with only up, down and strange quarks) in a magnetic field, many studies have been performed using SU(2) \cite{Braguta:2011hq,Luschevskaya:2012xd} and SU(3) \cite{Bali:2011qj, Hidaka:2012mz, Luschevskaya:2014lga} lattice gauge theory, AdS/QCD \cite{Callebaut:2011ab,Ammon:2011je,Callebaut:2013wba} as well as other phenomenological models \cite{Chernodub:2010qx,Chernodub:2011mc,Chernodub:2012zx,Simonov:2012if,Kojo:2012js,Li:2013aa,Andreichikov:2013zba,Orlovsky:2013wjd,Frasca:2013kka,Chernodub:2013uja,Liu:2014uwa,Taya:2014nha,Kawaguchi:2015gpt,Hattori:2015aki,Liu:2016vuw,Zhang:2016qrl}.
Some of them were devoted to discussions on $\rho$ meson condensation \cite{Chernodub:2010qx,Chernodub:2011mc}.
There have been a few more studies for light baryons
\cite{Andreichikov:2013pga,Haber:2014ula,Taya:2014nha,He:2015zca}.
It is known that the spontaneous chiral symmetry breaking ($\langle \bar{q} q \rangle$ condensate) in the QCD vacuum is enhanced by a magnetic field, which is the so-called magnetic catalysis of $\langle \bar{q} q \rangle$.
This effect has been studied using various QCD-like models \cite{Klevansky:1989vi,Suganuma:1990nn,Klimenko:1992ch,Gusynin:1994re,Gusynin:1994va,Gusynin:1994xp,Gusynin:1995nb,Hong:1996pv,Elias:1996zu,Shushpanov:1997sf,Ebert:1999ht,Agasian:1999sx,Kabat:2002er,Miransky:2002rp,Cohen:2007bt,Werbos:2007ym,Scherer:2012nn,Watson:2013ghq,Mueller:2014tea} as well as lattice QCD simulations \cite{Buividovich:2008wf, D'Elia:2010nq, Braguta:2010ej, Ilgenfritz:2012fw, Bali:2011qj, D'Elia:2011zu, Bali:2012zg, Bali:2013esa, Bruckmann:2013oba,Ilgenfritz:2013ara,Bali:2014kia,Endrodi:2015oba} (see Refs.~\cite{Shovkovy:2012zn,Miransky:2015ava} for recent reviews).
As temperature or density dependence of the $\langle \bar{q} q \rangle$ condensate can modify hadron properties, the contribution of the magnetic field dependence of the condensate to hadron properties is also one of the more attractive topics in hadron physics.

Properties of hadrons with heavy quarks (charm and bottom) in a magnetic field were also investigated by various approaches.
They include potential models \cite{Machado:2013rta,Alford:2013jva,Bonati:2015dka,Suzuki:2016kcs}, QCD sum rules \cite{Machado:2013yaa,Cho:2014exa,Cho:2014loa,Gubler:2015qok} and AdS/QCD \cite{Dudal:2014jfa}.
Among them, the constituent quark model is a suitable tool to investigate properties of heavy hadrons in an external field.
In such a model, there are mainly two physical effects which lead to the change of hadronic properties: (i) {\it mixing between different spin states} by the $-\bm{\mu}_i \cdot \bm{B}$ term and (ii) {\it modification of quark kinetic energy} by the $\bm{B} \times \bm{r}$ term.
In addition, for heavy-light meson systems such as $D$ and $B$ mesons, we expect two additional effects (iii) {\it Zeeman splitting} by the $-\bm{\mu}_i \cdot \bm{B}$ term and (iv) {\it magnetic field dependence of constituent light quark mass} induced by the magnetic catalysis.  
Furthermore, (v) {\it the anisotropy of the linear and Coulomb potential} between a quark and an antiquark manifests itself in a magnetic field \cite{Miransky:2002rp,Andreichikov:2012xe,Chernodub:2014uua,Bonati:2014ksa,Rougemont:2014efa,Simonov:2015yka}.
The latter effect was evaluated by recent lattice QCD simulations \cite{Bonati:2014ksa} and its correction in the potential model for quarkonia was investigated in Ref.~\cite{Bonati:2015dka}.

In this work, we will discuss the effects of (i)-(iv) by focusing on the neutral heavy meson systems.
The main purpose of this paper is a systematic understanding of heavy-meson spectra in a magnetic field,
while our previous work \cite{Suzuki:2016kcs} is devoted to the findings of the level structure and wave function deformation in charmonia.
To this end, we emphasize the importance of the comparison between charmonia and bottomonia,
which is characterized by their different masses and electric charge.
Furthermore, comparing quarkonia and heavy-light mesons is helpful to clarify the role of a light quark in heavy-meson systems under magnetic field.

It is not technically easy to solve the two-body Schr\"{o}dinger equation with a confining potential under strong magnetic field and to extract the anisotropic wave functions for ground and excited states simultaneously.
In Refs.~\cite{Alford:2013jva,Bonati:2015dka}, the finite differential time domain (FDTD) method is applied to obtain the eigenvalues and eigenstates for the ground states of quarkonia.
In our previous work \cite{Suzuki:2016kcs}, we proposed an alternative approach based on the variational method, which we call the cylindrical Gaussian expansion method (CGEM).
This method is an extension from the conventional Gaussian expansion method (GEM) \cite{Kamimura:1988zz,Hiyama:2003cu} mainly used for nuclear and atomic few-body systems.
The CGEM has the following nice properties: (i) it respects the symmetry of the Hamiltonian under constant magnetic field, (ii) it can deal fully with higher excited states and (iii) it reduces the computational cost substantially.

This paper is organized as follows.
In Sec. \ref{Sec_Formalism}, we construct a two-body Hamiltonian for heavy mesons in a magnetic field and develop the CGEM to solve the Schr\"{o}dinger equation in the cylindrical coordinate.
Numerical results obtained from this approach are shown in Sec. \ref{Sec_Results} and their physical pictures are discussed.
Section \ref{Sec_Conclusion and outlook} is devoted to our conclusion and outlook.

\section{Formalism} \label{Sec_Formalism}
\subsection{Quark model in a magnetic field}
Nonrelativistic two-body Hamiltonian of the constituent quark model in a magnetic field is written as \cite{Alford:2013jva}
\begin{equation}
H = \sum_{i=1}^2 \left[ \frac{1}{2m_i} \left( \bm{p}_i - q_i \bm{A} \right)^2 - \bm{\mu}_i \cdot \bm{B} +m_i \right] +V(r),
\end{equation}
where $m_i$, $\bm{p_i}$, $q_i$ and $\bm{\mu}_i$ are the constituent quark mass, the momentum of quarks, the quark electric charge and the quark magnetic moment, respectively.
For the vector potential, we choose the symmetric gauge $\bm{A}(\bm{r}_i) = \frac{1}{2} \bm{B} \times \bm{r}_i$.
We introduce the center of mass and relative coordinates, $\bm{R}=(m_1 \bm{r}_1 + m_2 \bm{r}_2)/M $ and $\bm{r} = \bm{r}_1 - \bm{r}_2$, where $M=m_1+m_2$ is the total mass of the two particles.
As a new conserved quantity, we define the pseudomomentum \cite{Avron:1978}: 
\begin{equation}
\hat{\bm{K}} = \sum_{i=1}^2 \left[ \bm{p}_i +\frac{1}{2} q_i \bm{B} \times \bm{r}_i  \right]
\end{equation}
whose commutation relation is given by $[\hat{K}_i, \hat{K}_j] = -i(q_1+q_2) \epsilon_{ijk} B_k$.
When the system is charge neutral, the components $\hat{K}_i$ commute with each other. 
By using the pseudomomentum, the total wave function for the Hamiltonian can be factorized as follows: 
\begin{equation}
\Phi(\bm{R},\bm{r}) = \exp \left[ i (\bm{K} -\frac{1}{2} q\bm{B} \times \bm{r}) \cdot \bm{R} \right] \Psi(\bm{r}).
\end{equation}

For a neutral system with $q_1=-q_2=q$, the total Hamiltonian can be reduced into
\begin{eqnarray}
H_\mathrm{rel} &=& \frac{\bm{K}^2}{2M} - \frac{q}{M}(\bm{K} \times \bm{B}) \cdot \bm{r} \nonumber\\
&& + \frac{q}{2} \left( \frac{1}{m_1} - \frac{1}{m_2} \right) \bm{B} \cdot( \bm{r} \times \bm{p}) \\
&& + \frac{\bm{p}^2}{2\mu} + \frac{q^2}{8\mu}( \bm{B} \times \bm{r})^2 + V(r) + \sum_{i=1}^2 [-\bm{\mu}_i \cdot \bm{B} + m_i], \nonumber
\end{eqnarray}
where $\bm{p}=\frac{m_2\bm{p_1}-m_1\bm{p_2}}{m_1+m_2}$, $\mu = m_1 m_2/M$ are the relative momentum and the reduced mass, respectively.
If the orbital angular momentum along $\bm{B}$ is zero, we have $\bm{B} \cdot (\bm{r} \times \bm{p}) = 0$. 
When we focus on a magnetic field along the $z$-axis, $\bm{B} =(0,0,B) $, $\bm{B} \times \bm{r} = B(-y,x,0)$ and $\bm{K} =(K_x,K_y,0) $, the Hamiltonian is reduced to, by using $p=-i \nabla$,
\begin{eqnarray}
H_\mathrm{rel} &=& \frac{\bm{K}^2}{2M} -\frac{\nabla^2}{2\mu} + \frac{q^2 B^2}{8\mu} \rho^2 + \frac{qB}{4\mu} K_x y - \frac{qB}{4\mu} K_y x  \nonumber\\
&& + V(r) + \sum_{i=1}^2 [-\bm{\mu}_i \cdot \bm{B} + m_i], \label{H_rel2}
\end{eqnarray}
where $\rho=\sqrt{x^2+y^2}$.
In this work, we investigate only the case of $\bm{K}=0$. In this situation, we can calculate the eigenvalues and eigenstates for mesons at rest. Then, the Hamiltonian maintains the rotational symmetry on the transverse $x$-$y$ plane and the reflection symmetry along the $z$-axis.

For the potential term in the model, we adopt the Cornell potential \cite{Eichten:1974af} with a spin-spin interaction:
\begin{eqnarray}
V(r) &=& \sigma r - \frac{A}{r} + \alpha (\bm{S}_1 \cdot \bm{S}_2)  e^{- \Lambda r^2} + C \nonumber\\
&=& \sigma \sqrt{\rho^2 + z^2} -\frac{A}{\sqrt{\rho^2 + z^2}} + \alpha (\bm{S}_1 \cdot \bm{S}_2) e^{- \Lambda (\rho^2 + z^2)} \nonumber\\
&& + C,
\end{eqnarray}
where $\bm{S}_1 \cdot \bm{S}_2 = -3/4$ and $1/4$ for the spin singlet and triplet, respectively.
Here, choosing the conventional Gaussian form for the spin-spin interaction leads to the analytic expressions of the matrix elements.

Let us now consider the contribution of the effects induced by the term $-\bm{\mu}_i \cdot \bm{B}$ \cite{Machado:2013rta,Alford:2013jva}, where the quark magnetic moment is given by $\bm{\mu}_i = g q_i \bm{S}_i/2m_i$ with the Land$\acute{\mathrm{e}}$ $g$-factor. We have assumed $g=2$.
As a result, for a neutral meson system, we obtain
\begin{eqnarray}
-(\bm{\mu}_1 + \bm{\mu}_2) \cdot \bm{B} &=& -\frac{gq}{2} \left( \frac{\bm{S}_1}{m_1} - \frac{\bm{S}_2}{m_2} \right) \cdot \bm{B} \nonumber\\
&=& -\frac{gq}{4} \left( \frac{\bm{\sigma}^1}{m_1} - \frac{\bm{\sigma}^2}{m_2} \right) \cdot \bm{B},
\end{eqnarray}
where $\bm{\sigma}^i$ is the Pauli matrix.
Then the eigenstates with the different spin quantum numbers, the singlet $| 00 \rangle =(| \uparrow \downarrow \rangle - | \downarrow \uparrow  \rangle) / \sqrt{2}$ and ``longitudinal" component $| 10 \rangle =(| \uparrow \downarrow \rangle + | \downarrow \uparrow  \rangle) / \sqrt{2}$ of the triplet, mix with each other through the off-diagonal matrix element:
\begin{equation}
\langle 10 | -(\bm{\mu}_1 + \bm{\mu}_2) \cdot \bm{B} |00 \rangle = -\frac{gqB}{4} \left( \frac{1}{m_1} + \frac{1}{m_2} \right). \label{mat_ele_1}
\end{equation}
To take into account the off-diagonal components of the matrix elements, Eq.~(\ref{mat_ele_1}), we have to solve a {\it coupled-channel} Schr\"{o}dinger equation.

On the other hand, ``transverse" components $| 1 \pm1 \rangle $ ($| \uparrow \uparrow \rangle$ and $| \downarrow \downarrow \rangle$) of the triplet cannot mix with other states and the off-diagonal components are zero.
For neutral systems with $m_1 \neq m_2$ (e.g. heavy-light mesons), the expectation values of these diagonal components are given by
\begin{equation}
\langle  1 \pm1 | -(\bm{\mu}_1 + \bm{\mu}_2) \cdot \bm{B} | 1 \pm1 \rangle = \mp \frac{gqB}{4} \left( \frac{1}{m_1} - \frac{1}{m_2} \right). \label{mat_ele_2}
\end{equation}
For systems with $m_1=m_2$ (quarkonia), these values become zero.
Then, the couplings between the quark magnetic moments and the magnetic field are completely canceled between the quark and the antiquark.

\subsection{Cylindrical Gaussian expansion method}
In this section, we review the cylindrical Gaussian expansion method.
After that, we introduce a coupled-channel basis to solve the coupled-channel Schr\"{o}dinger equation for the mixing between the spin singlet $| 00 \rangle$ and longitudinal component $| 10 \rangle$ of the spin triplet.

\subsubsection{Cylindrical Gaussian basis}
We have no analytic solution for the two-body Hamiltonian with the confinement potential and a magnetic field.
To solve the two-body Schr\"{o}dinger equation, we use the GEM based on the Rayleigh-Ritz variational method, which is a powerful tool in nuclear and atomic physics \cite{Kamimura:1988zz,Hiyama:2003cu}. 
In a magnetic field, a wave function should be expanded individually by the Gaussian basis on the transverse plane, $e^{- \beta_n \rho^2}$, and the basis along the longitudinal axis, $e^{- \gamma_n z^2}$, in the cylindrical coordinate $(\rho,z,\phi)$.
Such a cylindrical basis was first applied to the quark model in our previous work \cite{Suzuki:2016kcs} and a similar form was also successfully used for atomic systems such as hydrogen and helium in a magnetic field (e.g. Refs.~\cite{Aldrich1979,Becken:2001}).

The spatial part of the trial wave function for $l_z=0$ is given as follows \cite{Suzuki:2016kcs}:
\begin{eqnarray}
&& \Psi (\rho,z,\phi) = \sum_{n=1}^N C_n \Phi_n (\rho,z,\phi), \label{basis1} \\ 
&& \Phi_n (\rho,z,\phi) = N_n e^{- \beta_n \rho^2} e^{- \gamma_n z^2}, \label{basis2}
\end{eqnarray}
where $N$ and $C_n$ are the number of basis functions and the expansion coefficients.
$N_n$ is the normalization constant defined by $ \langle \Phi_n| \Phi_n \rangle =1$.
$\beta_n$ and $\gamma_n$ are Gaussian range parameters for the $\rho$ and $z$ directions, respectively.
It is empirically known that the best set of the range parameters, $\beta_1 \cdots \beta_N$ and $\gamma_1 \cdots \gamma_N$, are the geometric progressions:
\begin{eqnarray}
&& \beta_n = \frac{1}{\rho_n^2}, \ \ \ \rho_n = \rho_1 b^{n-1}, \label{geo1} \\ 
&& \gamma_n = \frac{1}{z_n^2}, \ \ \ z_n = z_1 c^{n-1}. \label{geo2}
\end{eqnarray}
Here, the four parameters, $\rho_1$, $\rho_N$, $z_1$ and $z_N$, will be optimized as the energy eigenvalue is minimized.
The analytic form of the matrix elements from the Hamiltonian (\ref{H_rel2}) and the wave function basis (\ref{basis1}) are summarized in Appendix \ref{App_mat}.
We checked the applicability of our numerical code by comparing our variational result with the analytic solution of the three-dimensional anisotropic harmonic oscillator.

\subsubsection{Coupled-channel basis for spin mixing}
Since the spin singlet $| 00 \rangle$ and longitudinal component $| 10 \rangle$ of the spin triplet mix with each other in a magnetic field as seen in Eq.~(\ref{mat_ele_1}), the wave function should take into account the coupled channel.
Then the basis function is given as a linear combination of the spin-0 and spin-1 components:
\begin{eqnarray}
&& \Psi (\rho,z,\phi) = \sum_{n=1}^N C^{S=0}_n \Phi^{S=0}_n (\rho,z,\phi) \nonumber \\
&&                 \hspace{43pt} + \sum_{n=N+1}^{2N} C^{S=1}_n \Phi^{S=1}_n (\rho,z,\phi), \label{basis3} \\
&& \Phi^{S=0}_n (\rho,z,\phi) = N^A_n e^{- \beta^A_n \rho^2} e^{- \gamma^A_n z^2}, \\
&& \Phi^{S=1}_n (\rho,z,\phi) = N^B_n e^{- \beta^B_n \rho^2} e^{- \gamma^B_n z^2},
\end{eqnarray}
where we have the four range parameters, $\beta^A_n$, $\gamma^A_n$, $\beta^B_n$ and $\gamma^B_n$.
These parameters are also expressed by the geometrical progression, so that we finally optimize the eight parameters.
A brief review of the generalized eigenvalue problem for the coupled-channel basis is shown in Appendix \ref{App_GEP}.

\subsection{Numerical setup} \label{subsec_parameter}
In this section, we set the parameters of our constituent quark model.
For parameters in charmonium systems, the charm-quark kinetic mass $m_c=1.7840 \, \mathrm{GeV}$, the Coulomb parameter $A=0.713$ and the string tension $\sqrt{\sigma} = 0.402 \, \mathrm{GeV}$ are determined from the equal-time $Q\bar{Q}$ Bethe-Salpeter amplitude in lattice QCD \cite{Kawanai:2015tga}.
For the spin-dependent potential given as $ \alpha \exp{(-\Lambda r^2)}$, we adopt $\Lambda=1.020 \, \mathrm{GeV}^2$ obtained from lattice QCD \cite{Kawanai:2011jt} by fitting a charmonium potential.
The remaining parameters, $\alpha=0.4778 \, \rm{GeV}$ and $C=-0.5693 \, \rm{GeV}$, are chosen to reproduce the experimental values of the masses of the ground states of $\eta_c$ and $J/\psi$.
We checked that these parameters can reproduce the masses of the ground and excited states of the charmonium well.

For bottomonium system, we have no information from lattice QCD.
Therefore, we fix $\sqrt{\sigma}$ on the same value as that of charmonium.
Furthermore, we put $C=0$ and tune the bottom-quark kinetic mass $m_b$, $A$ and $\alpha$ to reproduce the experimental masses of $\eta_b$, $\eta_b(2S)$ and $\Upsilon(1S)$.
For $D_s$ meson systems, we know the values of $m_s=0.656 \, \mathrm{GeV}$, $A=1.30 \, \mathrm{GeV}$ and $\sqrt{\sigma}=0.324 \, \mathrm{GeV}$ from the lattice QCD \cite{Kawanai:2015tga}.
Then, we tune only $\alpha$ and $C$ by matching our results with the experimental masses of $D_s$ and $D^\ast_s$ mesons.
For $D$, $B$ and $B_s$ meson systems, we fix $A$ and $\sqrt{\sigma}$ on the same values as those of $D_s$ meson.
The light-quark kinetic mass is fixed as $m_q=0.3 \, \mathrm{GeV}$ and only $\alpha$ and $C$ are tuned by the experimental masses of $D$--$D^\ast$, and $B$--$B^\ast$ and $B_s$--$B_s^\ast$ mesons.
Under these criteria, the resulting parameters are summarized in Tables \ref{paramter_list_M} and \ref{paramter_list_P}.
A list of the predicted meson masses is shown in Table \ref{paramter_list_com}.

\begin{table}[t!]
   \begin{center}
   \caption{Parameters for kinetic (constituent) quark masses.
$m_c$ for charmonium and $m_s$ for the $D_s$ meson were obtained from lattice QCD in Ref.~\cite{Kawanai:2015tga}.}
   \label{paramter_list_M}
  \begin{tabular}{c|c|c|c}
\hline \hline
$m_c \ [\mathrm{GeV}]$ & $m_b \ [\mathrm{GeV}]$ & $m_s \ [\mathrm{GeV}]$ &  $m_q \ [\mathrm{GeV}]$ \\
\hline
$1.784$ & $4.8078$ & $0.656$ &  $0.3$ \\
\hline \hline
   \end{tabular}
   \end{center}
\end{table}

\begin{table}[t!] 
   \begin{center}
   \caption{Parameters for potential terms.
$A$ and $\sqrt{\sigma}$ for charmonium and the $D_s$ meson were obtained from lattice QCD in Ref.~\cite{Kawanai:2015tga}.
$\Lambda$ for charmonium was estimated from the lattice QCD in Ref.~\cite{Kawanai:2011jt}. }
   \label{paramter_list_P}
   \begin{tabular}{l|c|c|c|c|c}
\hline \hline
Meson      & $A$     & $\sqrt{\sigma} \ [\mathrm{GeV}]$  & $\alpha \ [\mathrm{GeV}]$ & $\Lambda \ [\mathrm{GeV}^2]$ & $C \ [\mathrm{GeV}]$ \\
\hline
$c\bar{c}$          & $0.713$ & $0.402$  & $0.4778$  & $1.020$  & $-0.5693$ \\
$b\bar{b}$          & $0.531$ & $0.402$  & $0.1322$  & $1.020$  & $0$       \\
$D_s \, (c\bar{s})$ & $1.30$  & $0.324$  & $0.7954$  & $1.020$  & $-0.1765$ \\
$B_s \, (b\bar{s})$ & $1.30$  & $0.324$  & $0.2210$  & $1.020$  & $+0.2376$ \\
$D \, (c\bar{q})$   & $1.30$  & $0.324$  & $1.7812$  & $1.020$  & $-0.2419$ \\
$B \, (b\bar{q})$   & $1.30$  & $0.324$  & $0.5589$  & $1.020$  & $+0.1119$ \\
\hline \hline
   \end{tabular}
   \end{center}
\end{table}

\begin{table}[t!] 
   \begin{center}
   \caption{Comparison between meson masses predicted from parameters in Tables \ref{paramter_list_M} and \ref{paramter_list_P} and corresponding experimental values.}
   \label{paramter_list_com}
   \begin{tabular}{l|c|c|c}
\hline \hline
State              & Meson          & Our result $[\mathrm{GeV}]$  & Experiment $[\mathrm{GeV}]$ \cite{Agashe:2014kda} \\
\hline
$c\bar{c}$ $1^1S_0$ & $\eta_c(1S)$   &  2.984 &  2.984 \\
$c\bar{c}$ $1^3S_1$ & $J/\psi$       &  3.097 &  3.097 \\
$c\bar{c}$ $2^1S_0$ & $\eta_c(2S)$   &  3.669 &  3.639 \\
$c\bar{c}$ $2^3S_1$ & $\psi(2S)$     &  3.707 &  3.686 \\
\hline
$b\bar{b}$ $1^1S_0$ & $\eta_b(1S)$   &  9.398 &  9.398 \\
$b\bar{b}$ $1^3S_1$ & $\Upsilon(1S)$ &  9.460 &  9.460 \\
$b\bar{b}$ $2^1S_0$ & $\eta_b(2S)$   &  9.999 &  9.999 \\
$b\bar{b}$ $2^3S_1$ & $\Upsilon(2S)$ & 10.013 & 10.002 \\
$b\bar{b}$ $3^1S_0$ & $\eta_b(3S)$   & 10.330 &  ...   \\
$b\bar{b}$ $3^3S_1$ & $\Upsilon(2S)$ & 10.339 & 10.355 \\
\hline
$c\bar{s}$ $1^1S_0$ & $D_s$          & 1.968 &  1.968 \\
$c\bar{s}$ $1^3S_1$ & $D_s^\ast$     & 2.112 &  2.112 \\
\hline
$b\bar{s}$ $1^1S_0$ & $B_s$          & 5.367 &  5.367 \\
$b\bar{s}$ $1^3S_1$ & $B_s^\ast$     & 5.415 &  5.415 \\
\hline
$c\bar{q}$ $1^1S_0$ & $D$            & 1.870 &  1.870 \\
$c\bar{q}$ $1^3S_1$ & $D^\ast$       & 2.010 &  2.010 \\
\hline
$b\bar{q}$ $1^1S_0$ & $B$            & 5.279 &  5.279 \\
$b\bar{q}$ $1^3S_1$ & $B^\ast$       & 5.325 &  5.325 \\
\hline \hline
   \end{tabular}
   \end{center}
\end{table}

Here we comment on the possible magnetic-field dependence of those parameters.
As discussed in Refs.~\cite{Miransky:2002rp,Andreichikov:2012xe,Chernodub:2014uua,Bonati:2014ksa,Rougemont:2014efa,Simonov:2015yka}, the linear and Coulomb potential can have an anisotropy in a magnetic field, which leads to different $A$ and $\sqrt{\sigma}$ between the $\rho$ and $z$ directions.
Similarly, one can expect to appear $B$-dependences of the spin-spin interaction and constant term $C$.
In this work, although we do not consider those effects, such improvements would be interesting in the future.
Furthermore, the constituent light quark mass, $m_q$ and $m_s$, can be related to the chiral symmetry breaking in vacuum, so that the $B$-dependence of the chiral condensate can modify these masses.
Such effects in a heavy-light meson will be investigated in Sec. \ref{SubSec_MC}.

\section{Numerical results} \label{Sec_Results}

\begin{figure}[t!]
    \centering
    \includegraphics[clip, width=1.0\columnwidth]{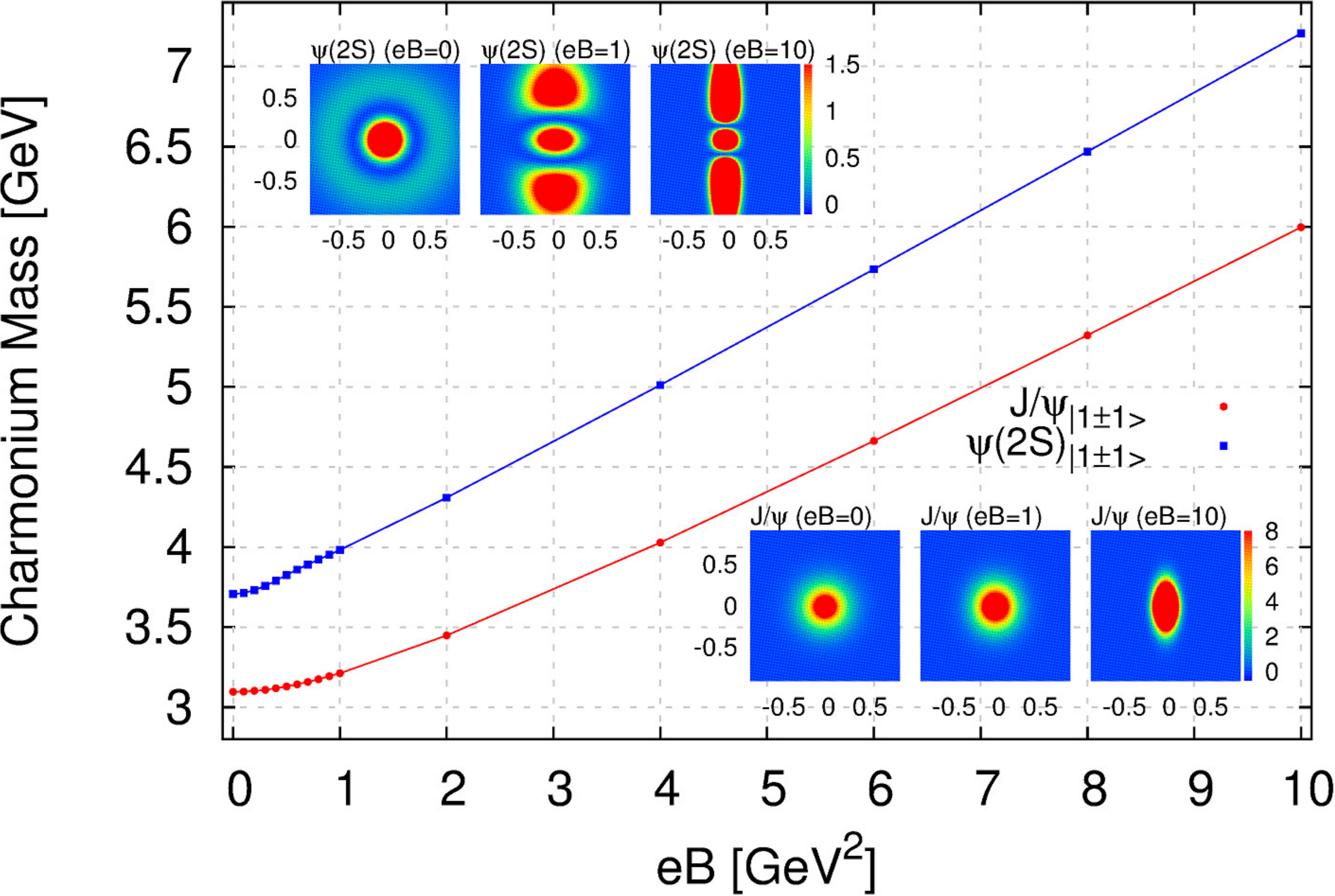}
    \caption{Masses of transverse $J/\psi, \psi(2S)$ in a magnetic field.
Densities of wave functions, defined by $|\Psi(\rho,z,\phi)|^2$, are plotted on the horizontal $x$ (or $y$)-vertical $z$ plane. }
    \label{Results_charmT_m}
\end{figure}

\begin{figure}[t!]
    \centering
    \includegraphics[clip, width=1.0\columnwidth]{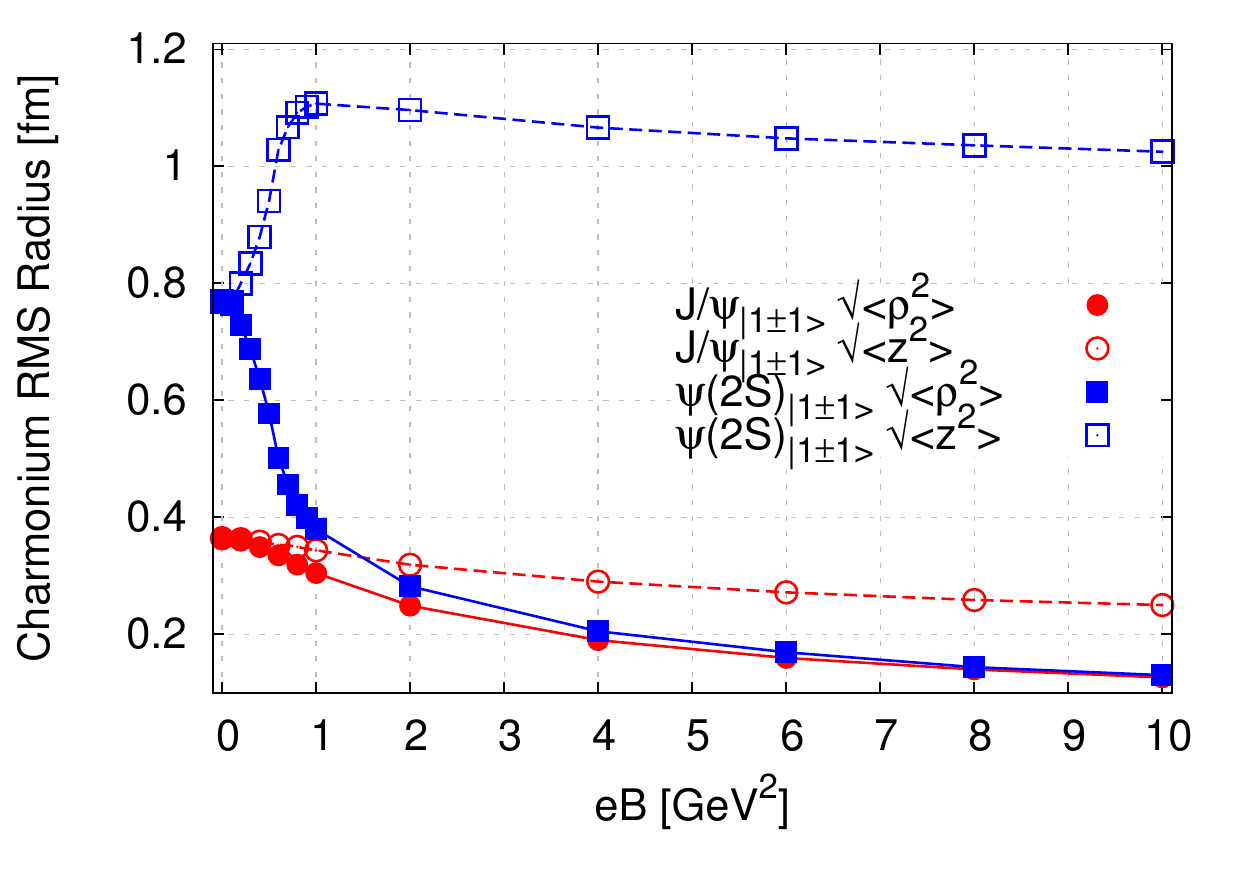}
    \caption{RMS radii of transverse $J/\psi, \psi(2S)$ in a magnetic field, defined by Eqs.~(\ref{RMS_rho}) and (\ref{RMS_z}).}
    \label{Results_charmT_r}
\end{figure}

\begin{figure*}[tb!]
    \centering
    \includegraphics[clip, width=17cm]{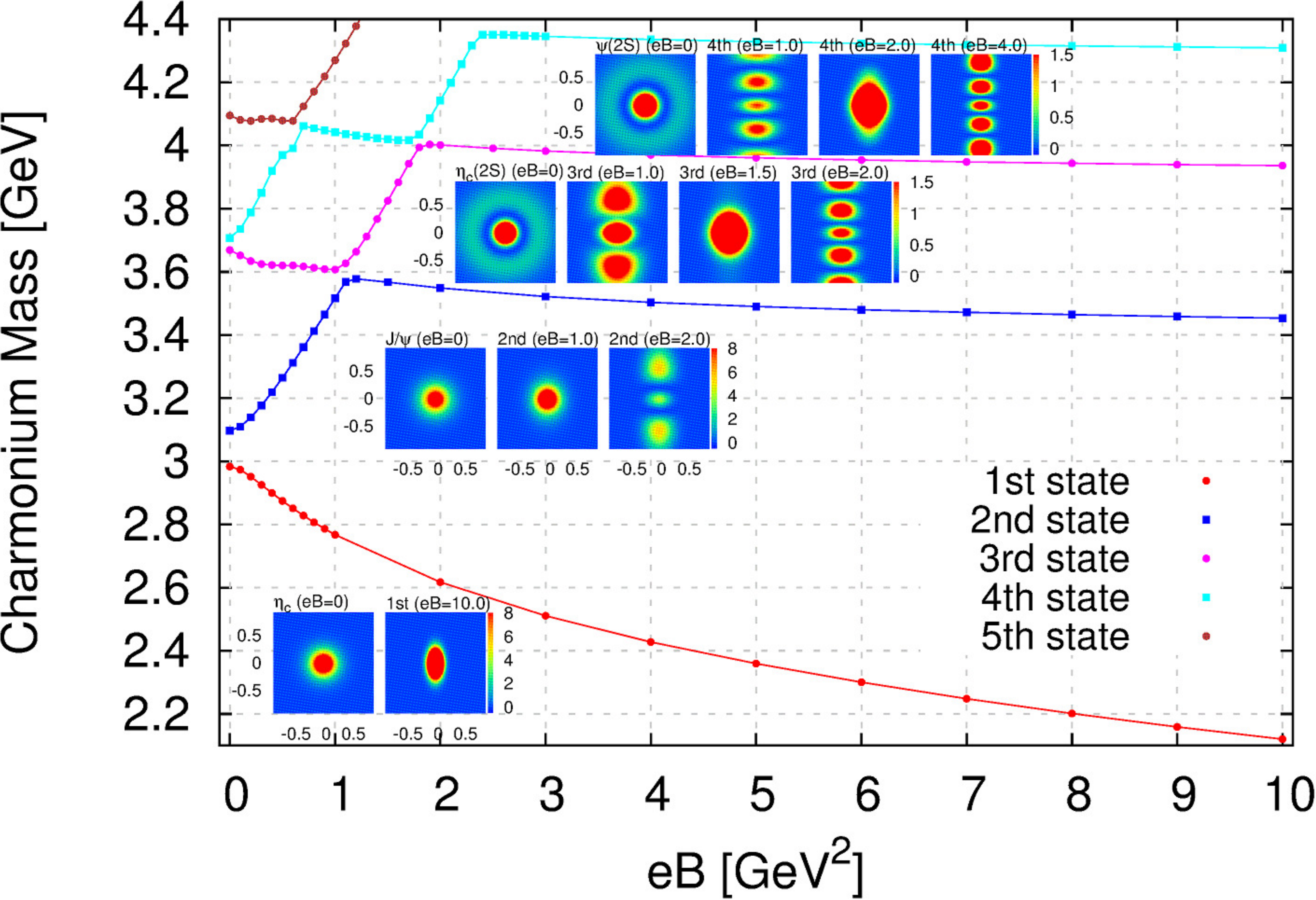}
    \caption{Masses of $\eta_c (1S,2S)$ and longitudinal $J/\psi$, $\psi(2S)$ in a magnetic field.}
    \label{Results_charmL_m}
\end{figure*}

\subsection{Charmonia}
The transverse vector charmonia [$J/\psi_{|1 \pm1 \rangle}$ and $\psi(2S)_{|1 \pm1 \rangle}$] in a magnetic field do not show mixing effects. 
The mass spectra and the shapes of the wave functions are shown in Fig.~\ref{Results_charmT_m}.
In a weak magnetic field, both the masses of the ground and excited states increase gradually.
We note that the mass shifts of the ground state estimated in weak magnetic fields agree with those obtained by using the same Hamiltonian and the FDTD method in Refs.~\cite{Alford:2013jva,Bonati:2015dka}.
We find that the excited state $\psi(2S)$ is more sensitive to a magnetic field than the ground state $J/\psi$.
For instance, the mass shifts at $eB=0.3 \, \mathrm{GeV}^2$ are $+12.4$ and $+50.8 \, \mathrm{MeV}$ for $J/\psi$ and $\psi(2S)$, respectively.
The reason is that the size of the wave function of $\psi(2S)$ is twice as large as that of the corresponding ground state, so that the expectation value of the $\rho^2$ term in Eq.~(\ref{H_rel2}) also becomes larger.
In a strong magnetic field, we see that the masses are linearly raised.
Such a mass shift can be quantitatively understood as follows.
The energy of a single constituent quark in a magnetic field is shifted by the {\it nonrelativistic} quark Landau level, $(n + 1/2)|q|B/m_c$, where $n$ is the discretized quantum number. 
The linear increase of the mass of a charmonium, as observed in our result, is consistent with the sum of such energy shifts for the two quarks inside the hadron.
Thus, in the strong fields, {\it the single-particle picture for the mass shifts is considered as a good approximation, even though it is not the rigorous two-body problem}.
We mention that, in Ref.~\cite{Taya:2014nha}, the masses of light mesons and baryons in a strong magnetic field were discussed from relativistic Landau levels of single particle quarks.

\begin{figure*}[tb!]
    \begin{minipage}[h]{1.0\columnwidth}
        \centering
        \includegraphics[clip, width=1.0\columnwidth]{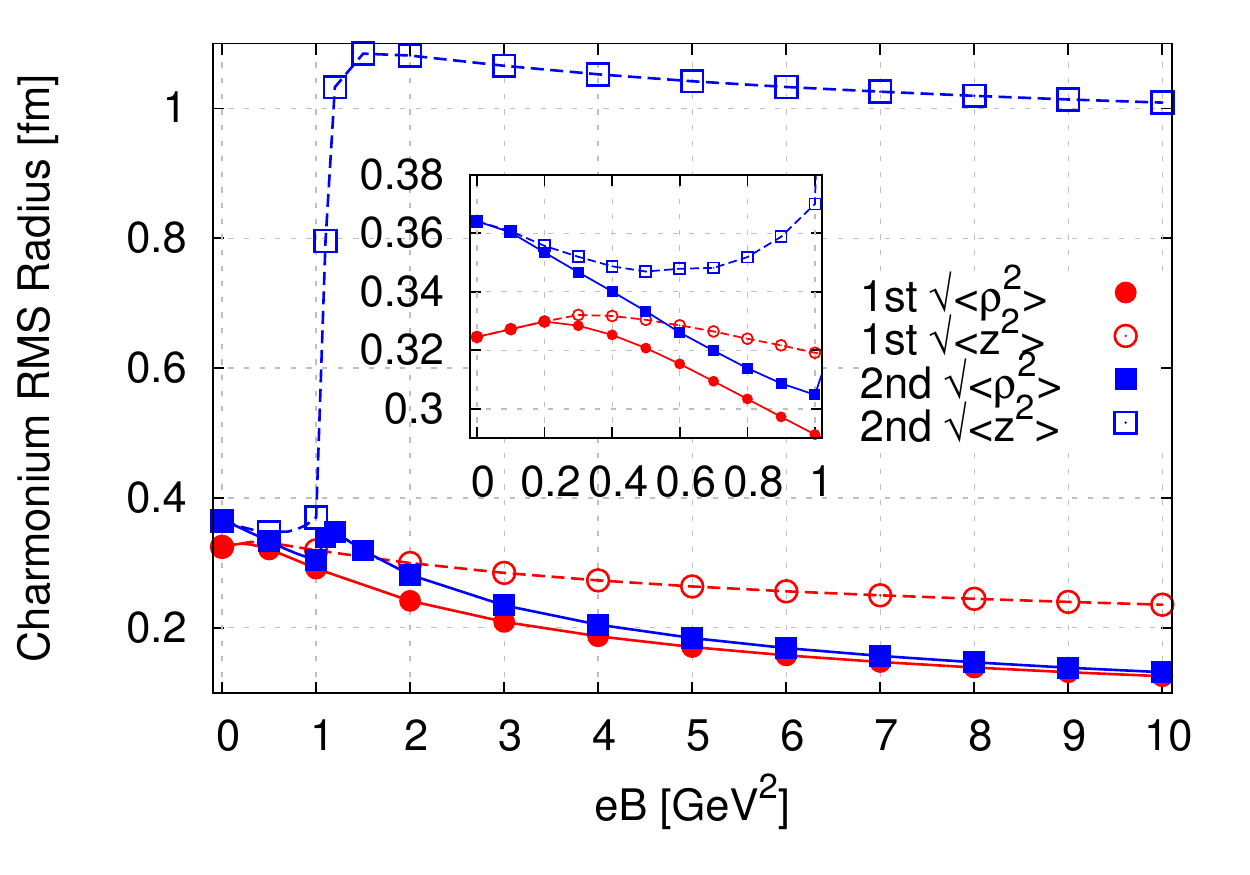}
    \end{minipage}%
    \begin{minipage}[h]{1.0\columnwidth}
        \centering
        \includegraphics[clip, width=1.0\columnwidth]{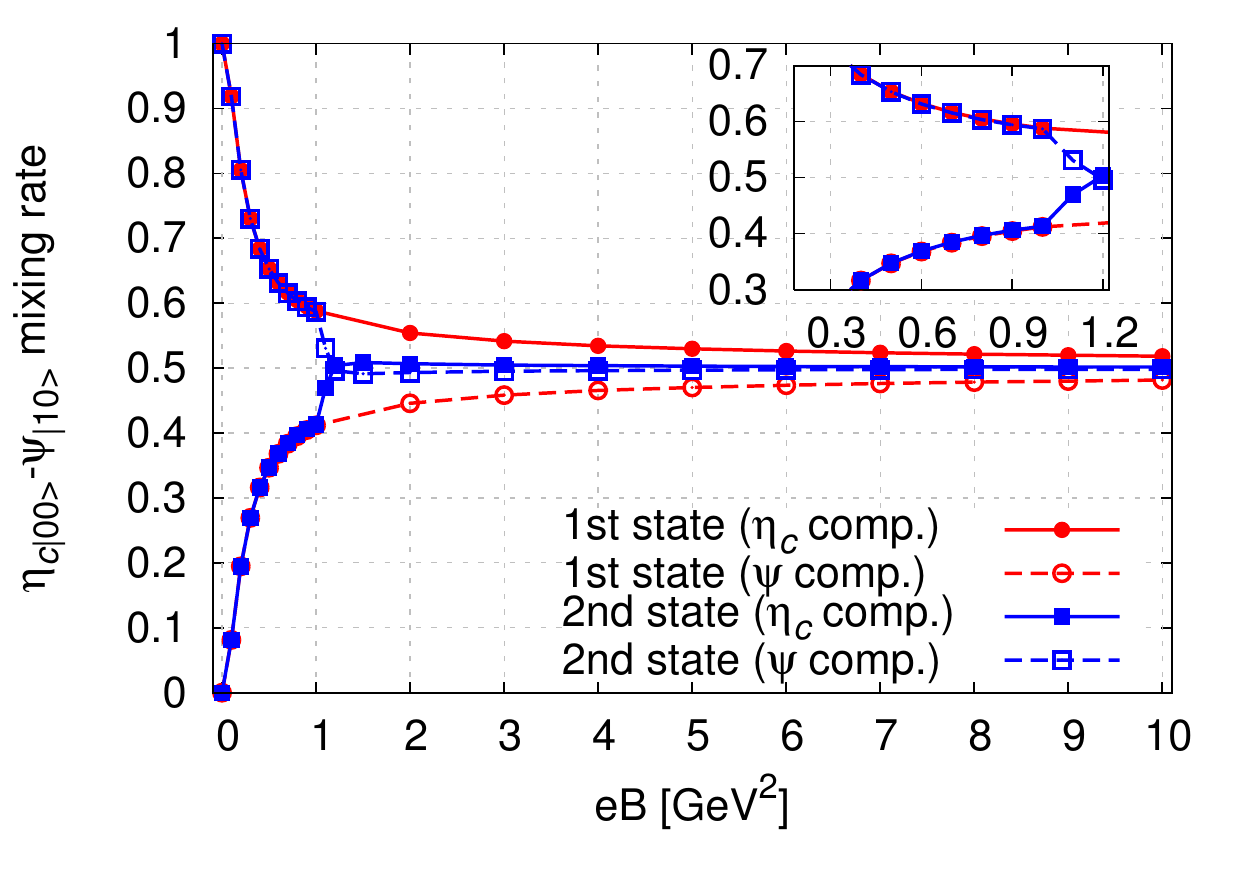}
    \end{minipage}
    \begin{minipage}[h]{1.0\columnwidth}
        \centering
        \includegraphics[clip, width=1.0\columnwidth]{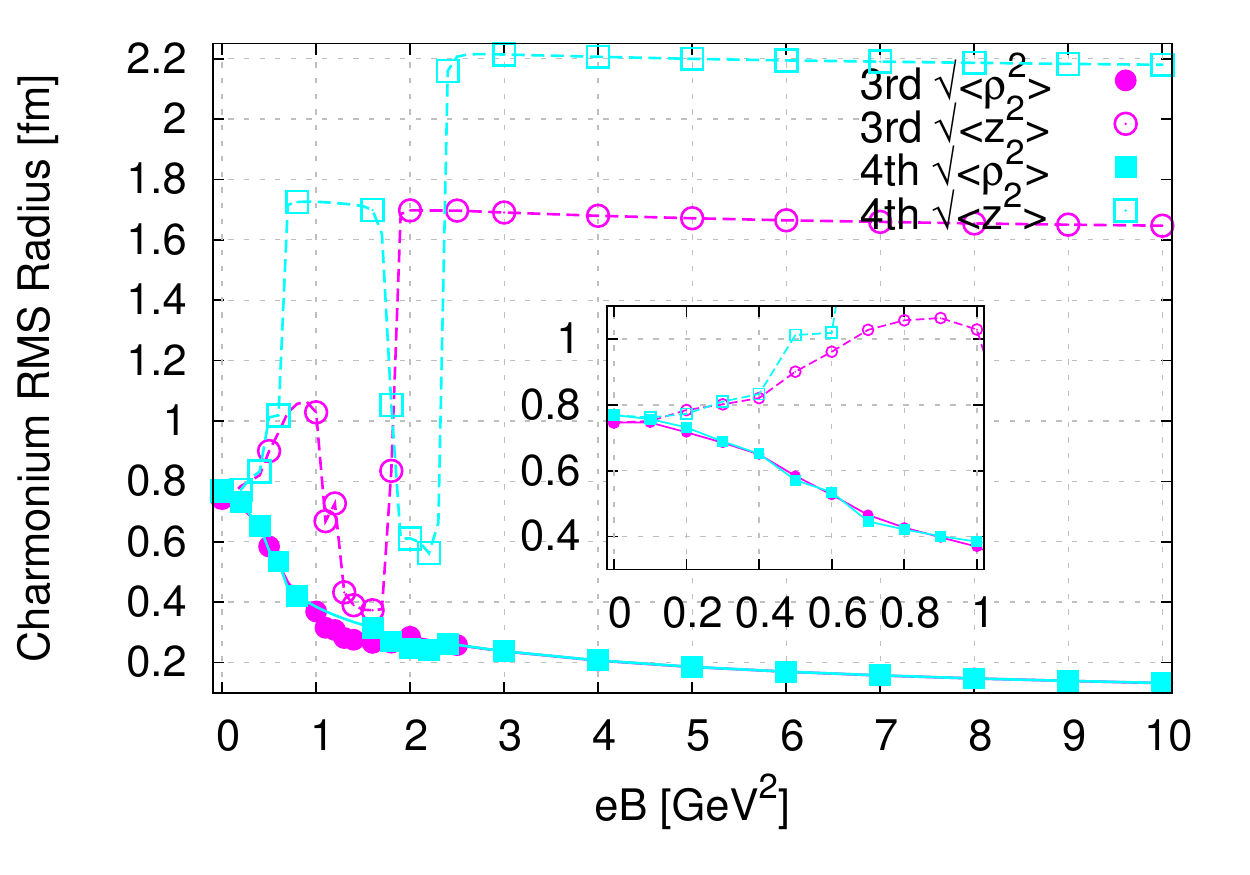}
    \end{minipage}%
    \begin{minipage}[h]{1.0\columnwidth}
        \centering
        \includegraphics[clip, width=1.0\columnwidth]{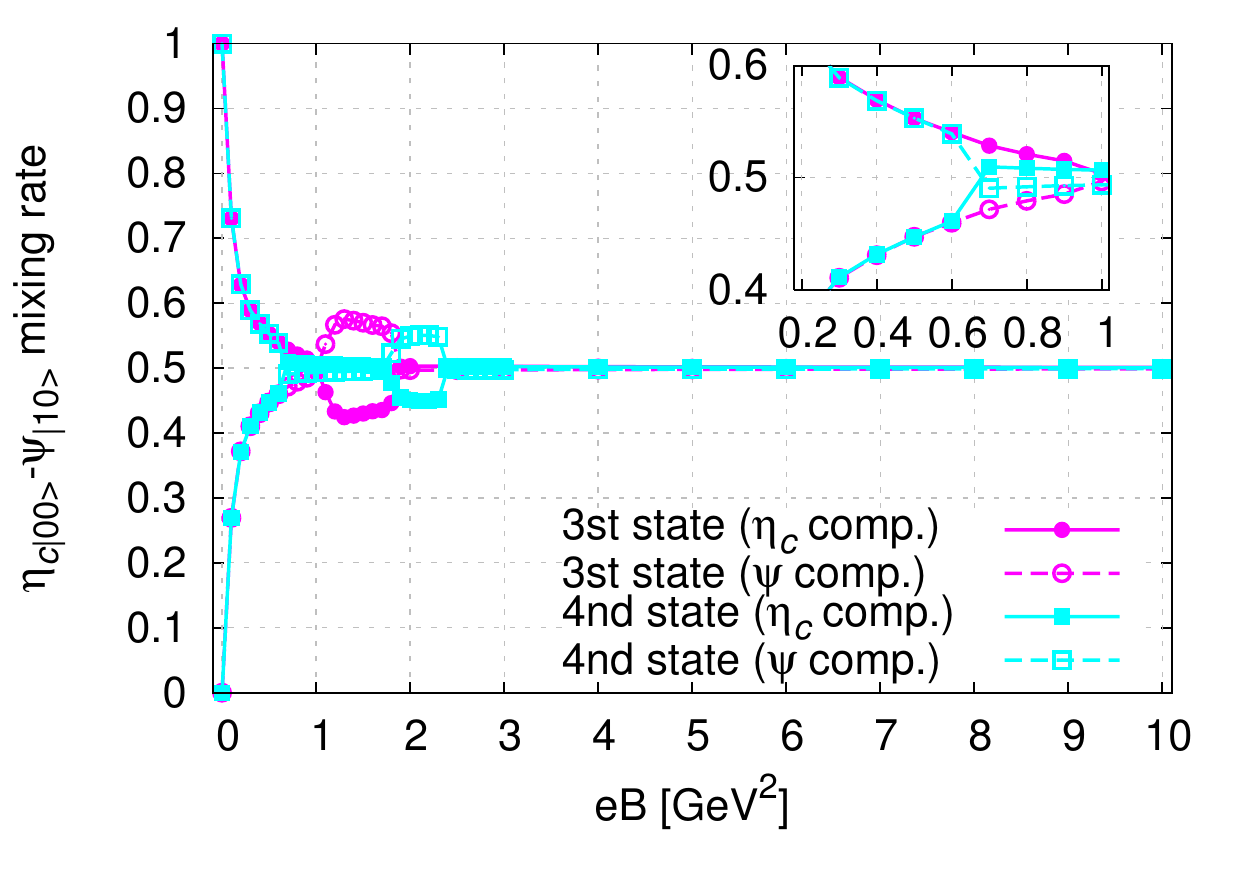}
    \end{minipage}
    \caption{RMS radii and mixing rates of $\eta_c (1S,2S)$ and longitudinal $J/\psi, \psi(2S)$ in a magnetic field.}
    \label{Results_charmL_r}
\end{figure*}

\begin{figure*}[t!]
    \centering
    \includegraphics[clip, width=17cm]{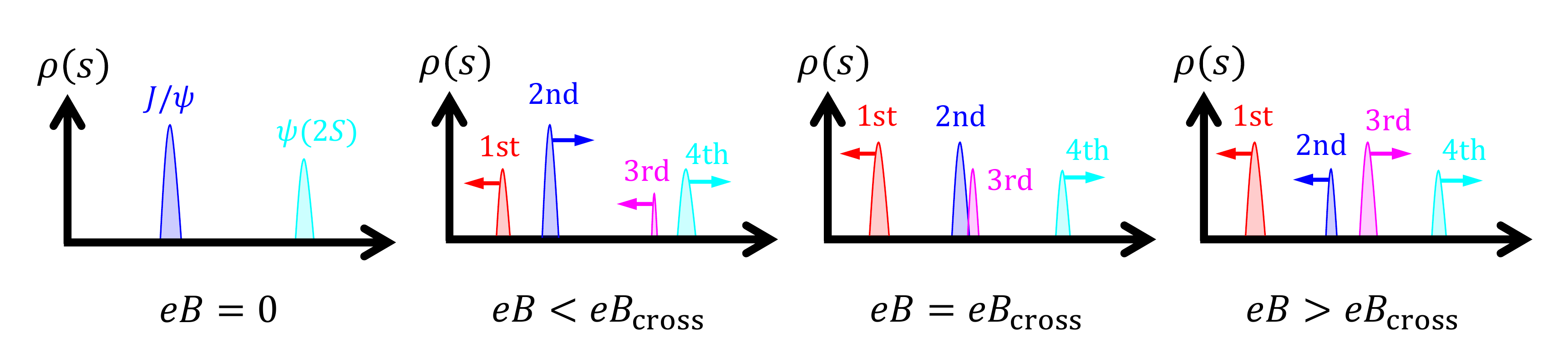}
    \caption{A schematic picture for an expected spectral function of longitudinal vector quarkonia in a magnetic field.
In zero field, there are only vector states corresponding to $J/\psi$ and $\psi(2S)$ in the spectral function.
In finite $eB$, the additional peaks such as the first and third states are newly induced.
Simultaneously the residues and positions of peaks can be modified.
By the level repulsion with the mixing partner, the second and third states approach each other.
$eB_\mathrm{cross}$ means the ''crossing magnetic field'' where the two peaks are located in the nearest positions.
After $eB_\mathrm{cross}$, the two peak positions start to separate.}
    \label{SPfunc}
\end{figure*}

In Fig.~\ref{Results_charmT_r}, we show the root-mean square (RMS) radii of the wave function of $J/\psi_{|1 \pm1 \rangle}$ and $\psi(2S)_{|1 \pm1 \rangle}$, calculated from the matrix elements in Appendix \ref{App_RMS}.
Here, we defined the RMS radii by
\begin{eqnarray}
&& \sqrt{\langle \rho^2 \rangle} \equiv \sqrt{(3/2) \langle \Psi | \rho^2 |\Psi \rangle}, \label{RMS_rho} \\
&& \sqrt{\langle z^2 \rangle} \equiv \sqrt{ 3 \langle \Psi | z^2 |\Psi \rangle}, \label{RMS_z} 
\end{eqnarray}
in the $\rho$- and $z$-directions, respectively.
These values in vacuum are the same as the usual spherical RMS radius $\sqrt{\langle r^2 \rangle} $: $\sqrt{\langle r^2 \rangle}_{B=0} = \sqrt{\langle \rho^2 \rangle}_{B=0} = \sqrt{\langle z^2 \rangle}_{B=0}$.
From this figure, we observe that both $\sqrt{\langle \rho^2 \rangle}$ and $\sqrt{\langle z^2 \rangle}$ of $J/\psi_{|1 \pm1 \rangle}$ decrease gradually.
The shrinkage on the $\rho$-plane is induced by the $\rho^2$ term in Eq.~(\ref{H_rel2}) as expected naively.
In the case of $\psi(2S)_{|1 \pm1 \rangle}$, $\sqrt{\langle \rho^2 \rangle}$ decreases gradually in all the regions of the magnetic field while $\sqrt{\langle z^2 \rangle}$ increases in weak magnetic fields.
At about $eB=1.0 \, \mathrm{GeV}$, the enlargement of the wave function in the $z$-direction saturates, and, after that, it begins to decrease little by little.
Then, in the wave function of a $2S$ state, only the excitation in the $z$-direction survives, which means that it behaves as a quasi-one-dimensional object.

The results for pseudoscalar [$\eta_c(1S)$ and $\eta_c(2S)$] and longitudinal vector [$J/\psi_{|1 0 \rangle}$ and $\psi(2S)_{|1 0 \rangle}$] charmonia are shown in Figs.~\ref{Results_charmL_m} and \ref{Results_charmL_r}.
In these hadrons in a magnetic field, spin-0 ($\eta_c$) and spin-1 ($\psi$) components are mixed with each other by the $-\bm{\mu}_i \cdot \bm{B}$ term in Eq.~(\ref{H_rel2}), so that we could no longer call these states $\eta_c$ or $J/\psi$ .
Therefore, we will label these states as {\it first}, {\it second}, {\it third}, and {\it fourth states}.

For the {\it first} state starting from $\eta_c(1S)$ in vacuum, the mixing partners include all spin-1 excited states [$\psi(2S)$, $\psi(3S)$ ...] as well as the ground state $J/\psi$.
$\eta_c(1S)$ is expected to be predominantly mixed with $J/\psi$ and its mass decreases gradually by the level repulsion with increasing magnetic field.
Such predominance of the mixing partners is also shown as the agreement between the first and second states in the figure of the mixing rates as shown in the upper right of Fig.~\ref{Results_charmL_r}.
One of the interesting points is that the RMS radii in both $\rho$- and $z$-directions, for the first state in weak magnetic fields up to $eB=0.2 \, \mathrm{GeV}^2$, seem to increase slightly, as shown in the upper left of Fig.~\ref{Results_charmL_r}.
The behavior might come from the contamination of the $J/\psi$ components with the larger RMS radius.
The radii of the second state are in contrast decreasing as $B$ grows.
The {\it second} state starting from $J/\psi$ in vacuum is also expected to be mixed with $\eta_c(1S)$ in a weak magnetic field.
Its mass approaches the third state like $\eta_c(2S)$ at a ``crossing magnetic field'' located in about $eB=1.0$-$1.1 \, \mathrm{GeV}^2$.
Such a crossing point (strictly, avoided crossing) can also be seen as a jump in the figure of the RMS radii and the mixing rates.
In this region, the dominance of the $\psi$ component, characterized by the mixing rate more than $50\%$, is reversed, and the rate approaches almost $50\%$.
After the crossing, the wave function is converted to the $2S$-like shape with one node, and the mass of the second state begins to decrease slowly.
Note that, as with the case of transverse $J/\psi$, the mass shifts of the first and second states in the weak magnetic fields obtained by the previous works in Refs.~\cite{Alford:2013jva,Bonati:2015dka} are consistent with our results.
We also checked the comparison with the result from a hadron effective Lagrangian suggested in Refs.~\cite{Cho:2014exa,Cho:2014loa} in Appendix~\ref{App_EFT}.

Furthermore, the {\it third} state starting from $\eta_c(2S)$ can be expected to mix with the $\psi(2S)$ state in a weak magnetic field. 
An important point is that the mass shifts of the excited states in the weak magnetic field are more sensitive to $eB$ than that of the ground state.
For instance, the obtained mass shifts of the first, second, third and fourth states at $eB=0.1 \, \mathrm{GeV}^2$ are $-9.9$, $+12.4$, $-16.6$ and $+28.8 \, \mathrm{MeV}$, respectively.
One of the reasons is that the hyperfine splitting in vacuum is smaller so that their mixing in a magnetic field becomes stronger: $\Delta m_{J/\psi-\eta_c}=113\mathrm{MeV}$ for the ground state and $\Delta m_{\psi(2S)-\eta_c(2S)}=47\mathrm{MeV}$ for the excited state \cite{Agashe:2014kda}.
Such a tendency can be easily inferred from a simplified two-level model as discussed in Refs.~\cite{Alford:2013jva,Cho:2014exa,Cho:2014loa} (also see Appendix.~\ref{App_EFT}).
Thus the mixing in the weak magnetic fields quantitatively agrees with the behavior in the two-level system while the behavior in intermediate magnetic fields becomes more complicated because of the mixing between various states.
Then the qualitative behavior of RMS radii is similar to the case of transverse $J/\psi$ as shown in Fig.~\ref{Results_charmT_r}.
At $eB=1.0$-$1.1 \, \mathrm{GeV}^2$, the third state approaches the second one like $J/\psi$.
After the crossing, the mixing rate of the state shows a $\psi$-component dominance, which means that the state is predominantly mixed with the first state like $\eta_c(1S)$.
After that, the state approaches the fourth one like $\eta_c(3S)$ at $eB=1.8$-$1.9 \, \mathrm{GeV}^2$ and the wave function changes to the $3S$-like shape with two nodes.
Finally, we also find that the {\it fourth} state starting from $\psi(2S)$ shows similar behavior, which approaches the fifth state like $\eta_c(3S)$ at $eB=0.6$-$0.7 \, \mathrm{GeV}^2$, the third state like $J/\psi$ and the fifth state like $\eta_c(4S)$ at $eB=2.4$-$2.5 \, \mathrm{GeV}^2$.

\begin{figure}[t!]
    \centering
    \includegraphics[clip, width=1.0\columnwidth]{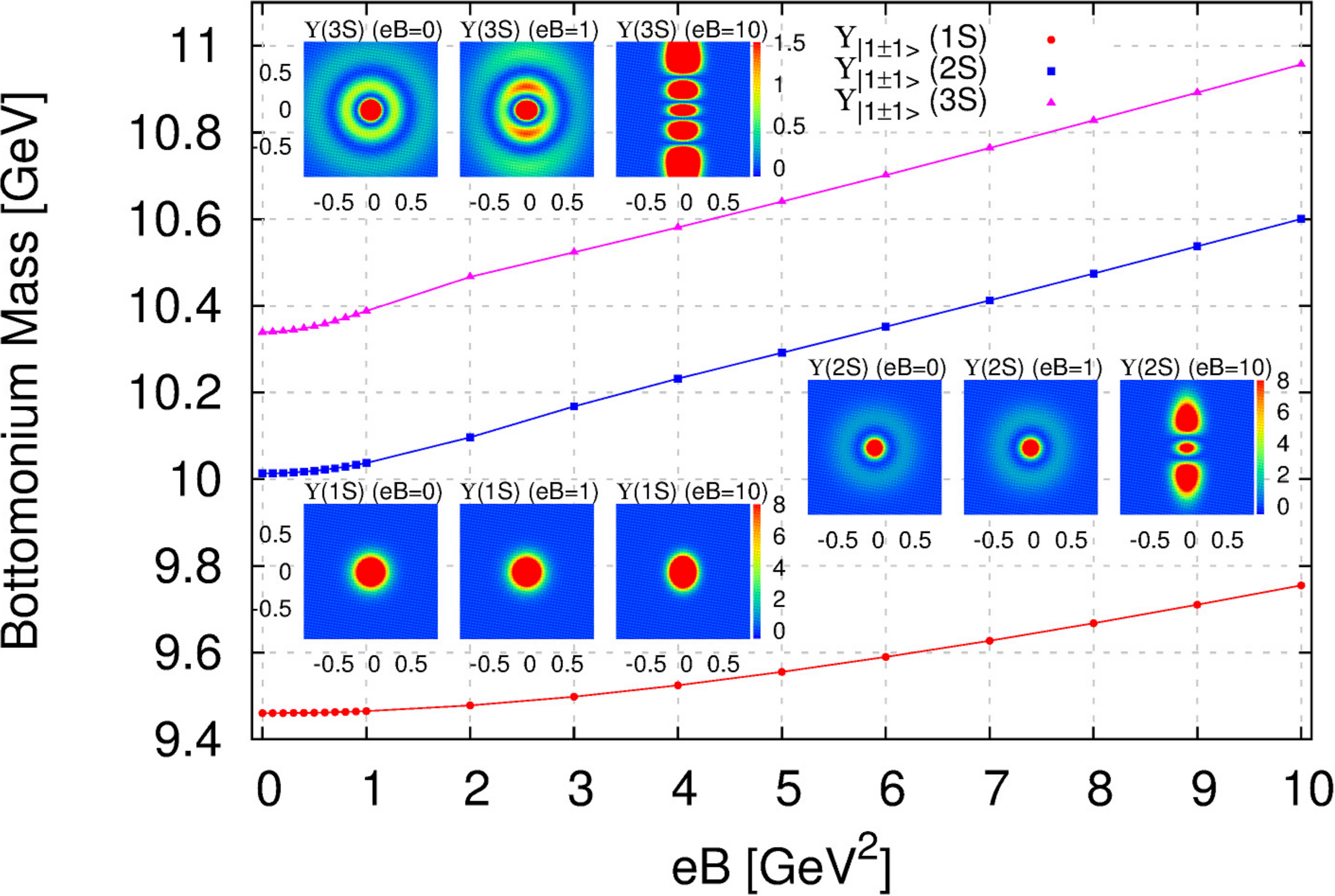}
    \caption{Masses of transverse $\Upsilon (1S,2S,3S)$ in a magnetic field.}
    \label{Results_bottomT_m}
\end{figure}

\begin{figure}[t!]
    \centering
    \includegraphics[clip, width=1.0\columnwidth]{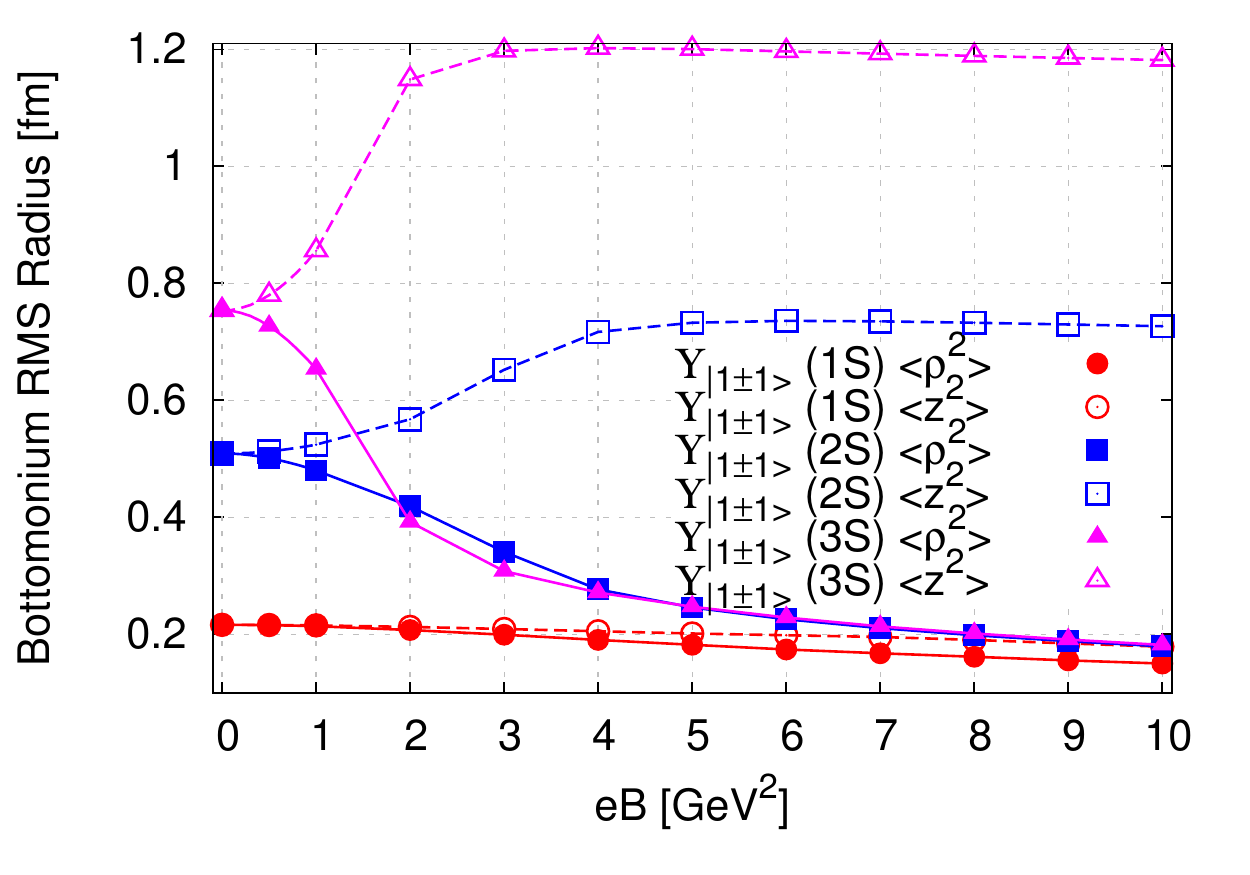}
    \caption{RMS radii of transverse $\Upsilon (1S,2S,3S)$ in a magnetic field.}
    \label{Results_bottomT_r}
\end{figure}

\begin{figure*}[t!]
    \centering
    \includegraphics[clip,width=17cm]{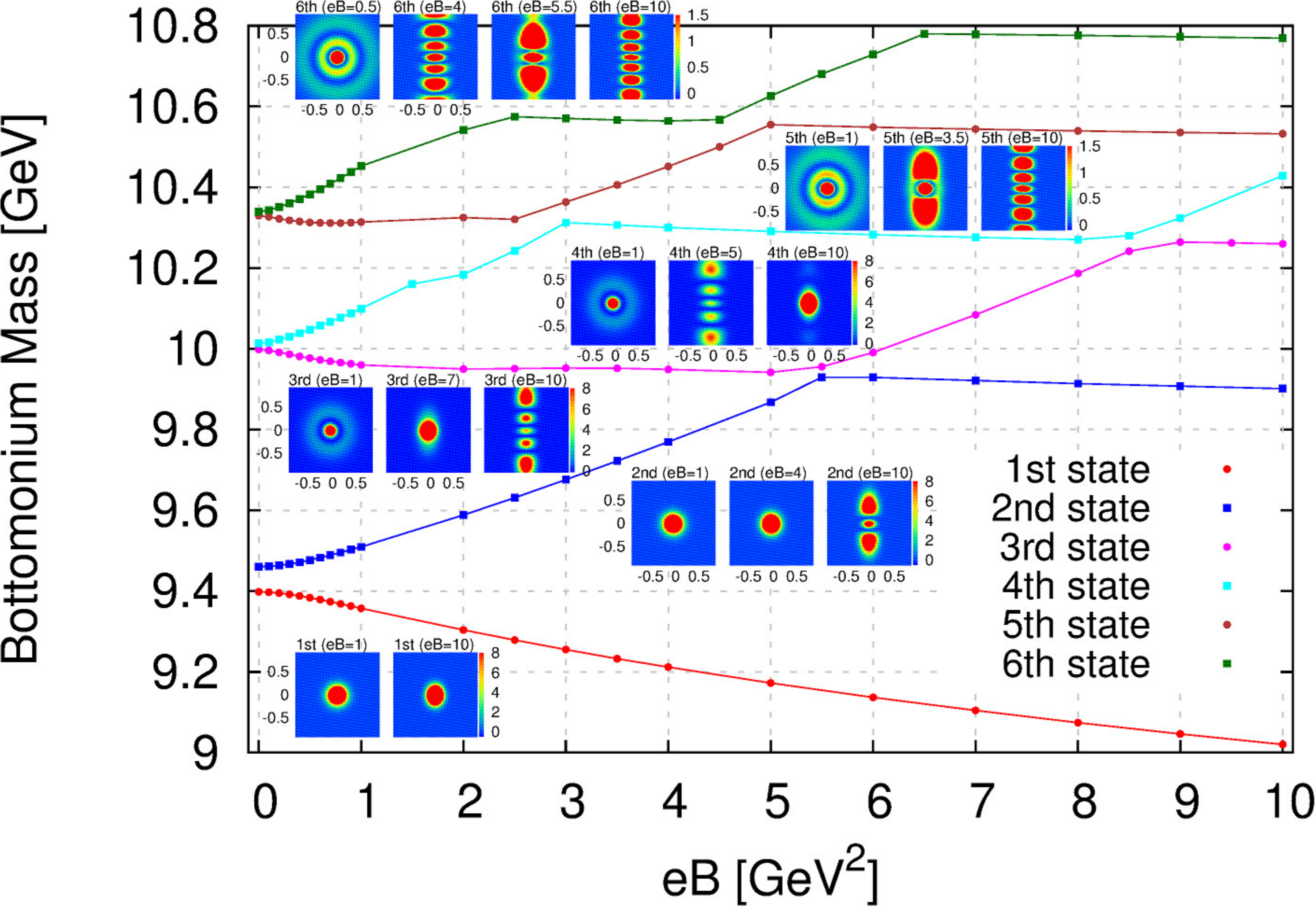}
    \caption{Masses of $\eta_b (1S,2S,3S)$ and longitudinal $\Upsilon (1S,2S,3S)$ in a magnetic field.}
    \label{Results_bottomL_m}
\end{figure*}

Here, we emphasize that the quantitative determination of crossing magnetic field would be important to compare our prediction with observables from other theoretical approaches or experiments.
When we obtain hadron masses from QCD sum rules and lattice QCD simulations, we may evaluate a spectral function from the corresponding hadronic current-current correlator.
For example, in Fig.~\ref{SPfunc}, we show expected spectral functions of the longitudinal vector quarkonia. 
Since $J/\psi$ mixes with $\eta_c$ components in a magnetic field, the spectral function obtained from the vector-vector current correlator in a magnetic field can have not only the ``original" $J/\psi$ and $\psi(2S)$ peaks (and continuum) but also ``additional" peaks such as $\eta_c$ and $\eta_c(2S)$ \cite{Cho:2014exa,Cho:2014loa}.
Then the level crossing in a magnetic field means that the positions of the two peaks in the spectral functions approach each other.
After that, the two peaks are expected not to slip thorough but rather to begin to repel each other by the avoided crossing. 
Although it might be usually difficult to extract the higher-energy structures except for the ground state from QCD sum rules and lattice QCD simulations, phenomenological estimation as constructed by Refs.~\cite{Cho:2014exa,Cho:2014loa} and some numerical techniques such as the maximum entropy method \cite{Asakawa:2000tr,Gubler:2010cf,Araki:2014qya} could be useful to investigate excited states and crossing structures in hadronic spectral functions under finite magnetic field.
Furthermore, the spectral function for vector channels can be related to the experimental dilepton spectra (e.g. $J/\psi \to e^+ e^-$ in vacuum) and the modifications of such observables by mixing (crossing) effect under magnetic field would be more interesting.  

\subsection{Bottomonia}

\begin{figure*}[t!]
    \begin{minipage}[h]{1.0\columnwidth}
        \centering
        \includegraphics[clip, width=1.0\columnwidth]{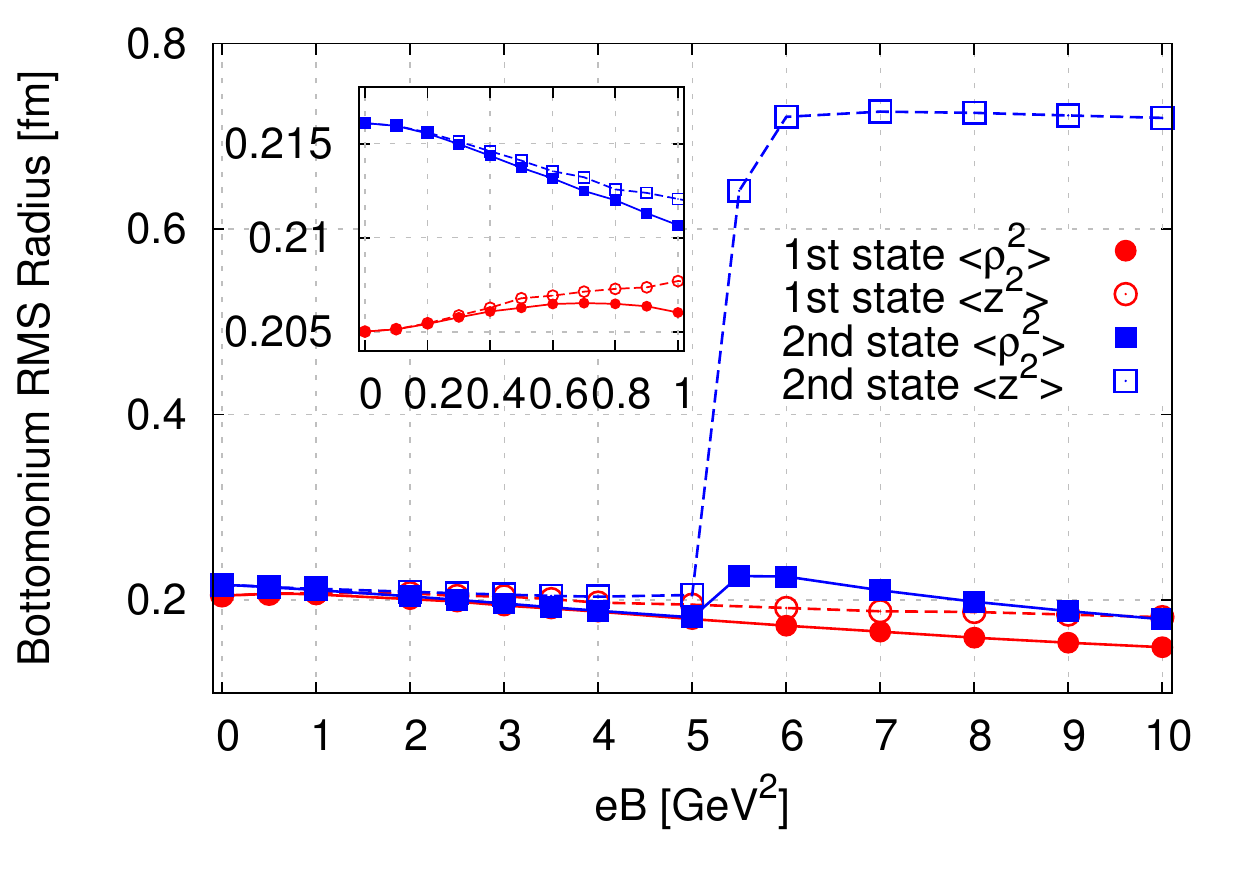}
    \end{minipage}%
    \begin{minipage}[h]{1.0\columnwidth}
        \centering
        \includegraphics[clip, width=1.0\columnwidth]{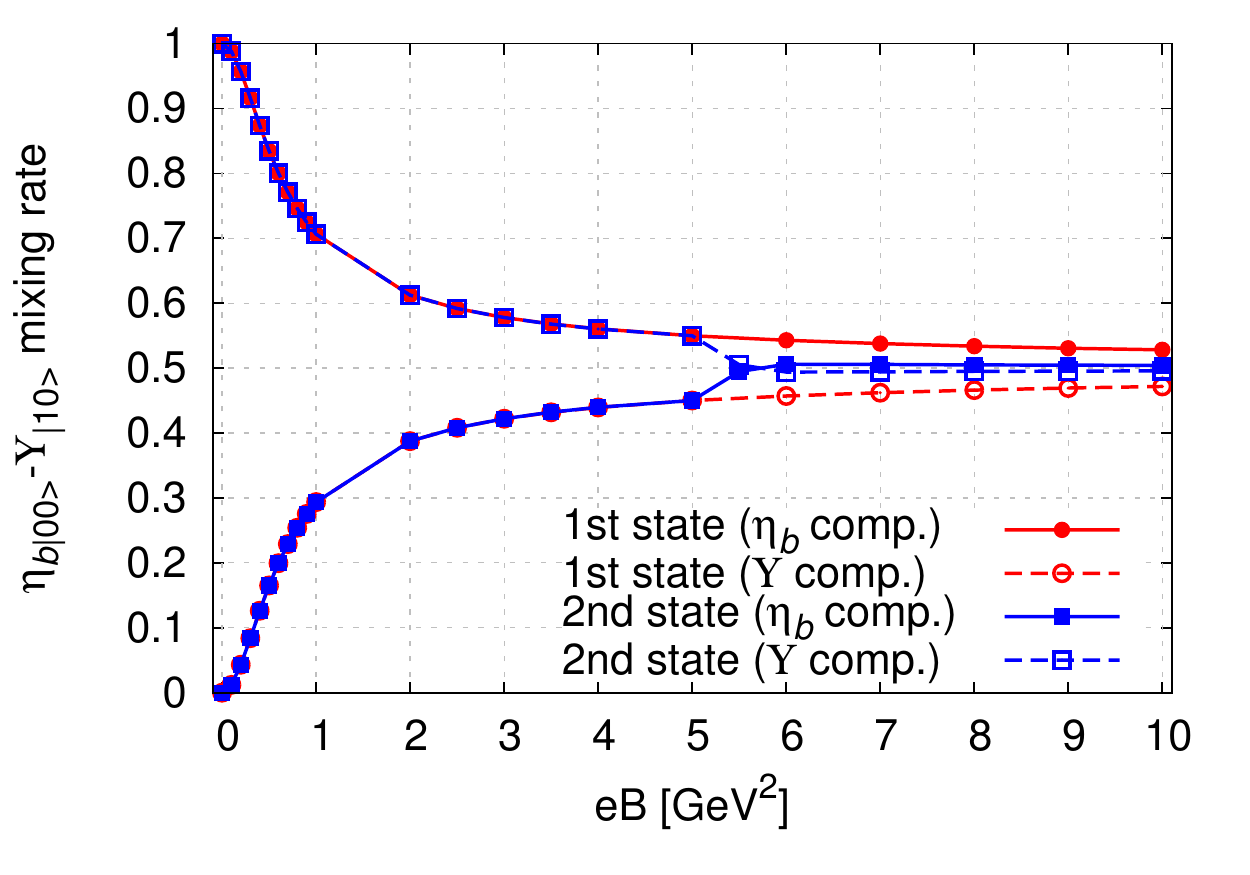}
    \end{minipage}
    \begin{minipage}[h]{1.0\columnwidth}
        \centering
        \includegraphics[clip, width=1.0\columnwidth]{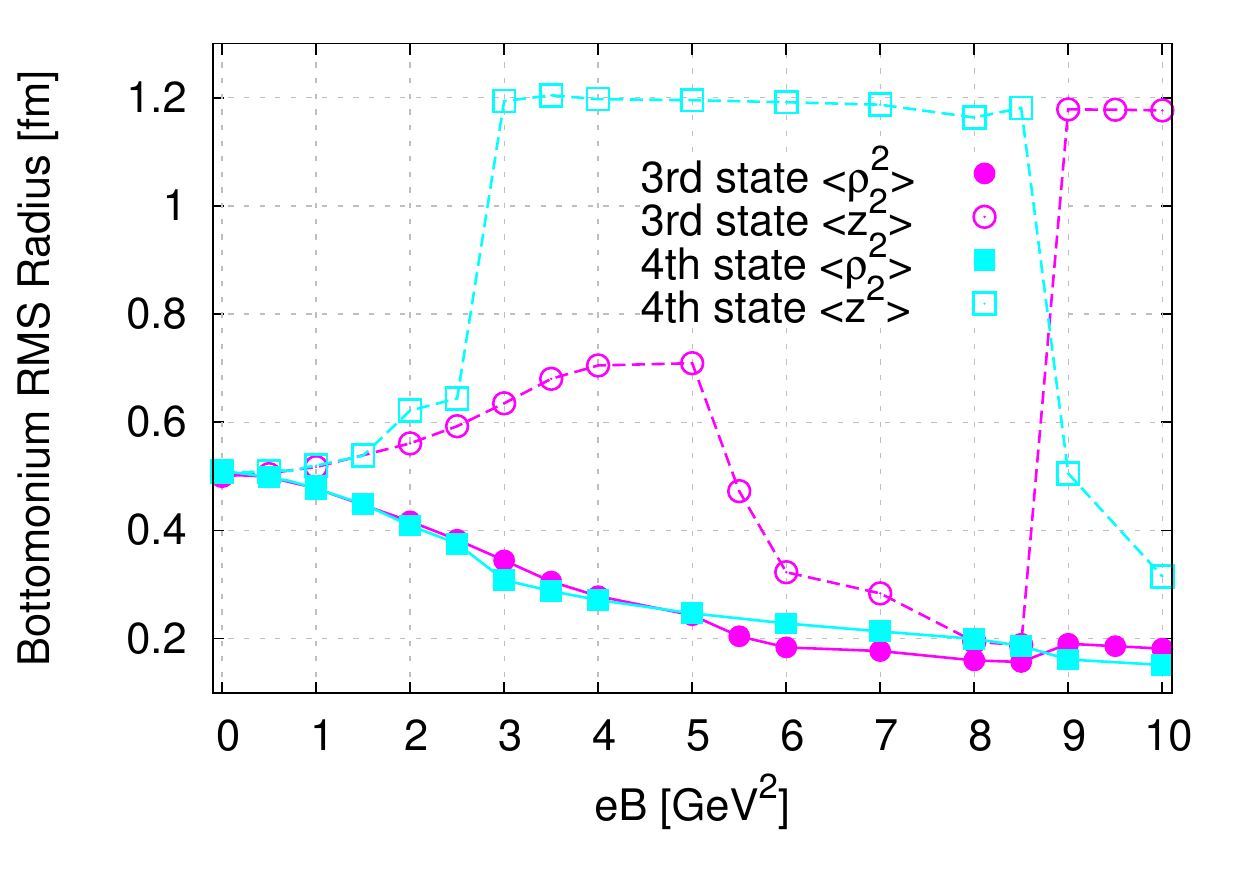}
    \end{minipage}%
    \begin{minipage}[h]{1.0\columnwidth}
        \centering
        \includegraphics[clip, width=1.0\columnwidth]{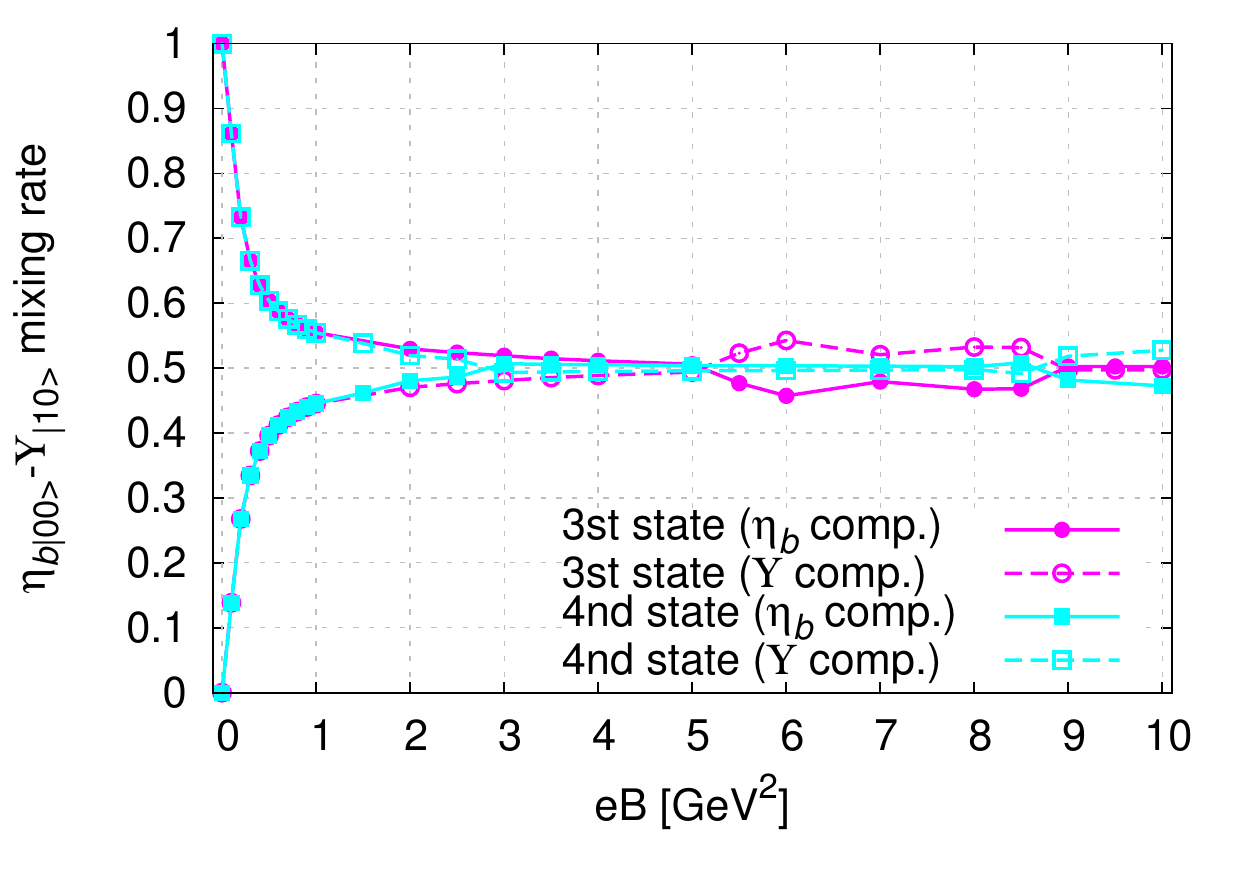}
    \end{minipage}
        \begin{minipage}[h]{1.0\columnwidth}
        \centering
        \includegraphics[clip, width=1.0\columnwidth]{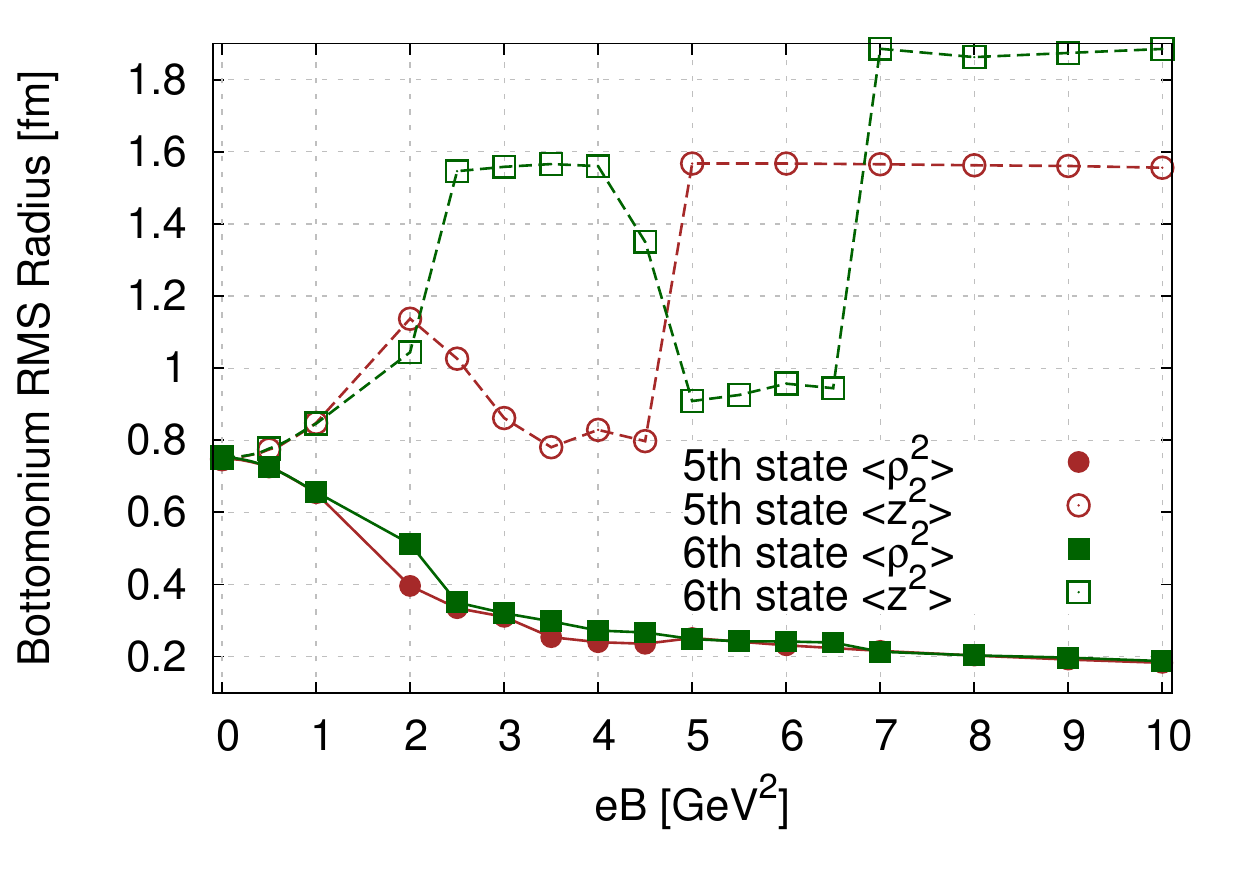}
    \end{minipage}%
    \begin{minipage}[h]{1.0\columnwidth}
        \centering
        \includegraphics[clip, width=1.0\columnwidth]{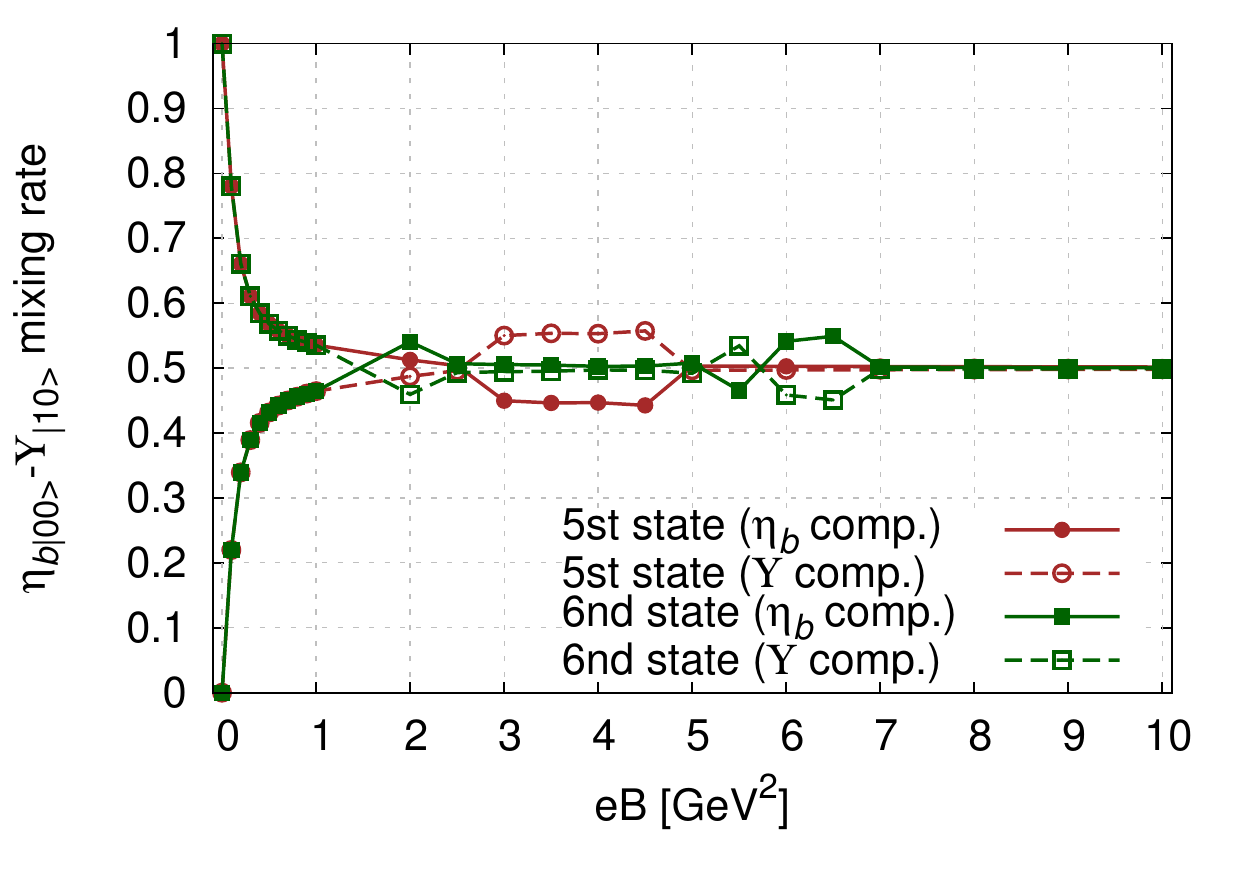}
    \end{minipage}
    \caption{RMS radii and mixing rates of $\eta_b (1S,2S,3S)$ and longitudinal $\Upsilon (1S,2S,3S)$ in a magnetic field.}
    \label{Results_bottomL_r}
\end{figure*}

Next, we move to the discussions on bottomonia.
For pseudoscalar and vector bottomonia, there are six states below the $B\bar{B}$ threshold located at about $10.56 \, \mathrm{GeV}$: $\eta_b(1S)$, $\Upsilon(1S)$, $\eta_b(2S)$, $\Upsilon(2S)$, $\eta_b(3S)$, and $\Upsilon(3S)$.
The main differences between charmonia and bottomonia are (i) quark mass and (ii) quark electric charge.
For (i), since bottom quarks are heavier than charm quarks, the kinetic energies of the heavy quarks moving inside a quarkonia are suppressed and the sizes of the wave functions tend to be smaller than those of charmonia.
Furthermore, for (ii), the electric charge of bottom quarks is $|q|=(1/3)|e|$ while that of charm quarks is $|q|=(2/3)|e|$.
This means that bottomonia are less sensitive to a magnetic field than charmonia, as seen in the $(q^2 B^2/8\mu) \rho^2$ term in Eq.~(\ref{H_rel2}).
From the definition of the magnetic moment, $\bm{\mu}_i = g q_i \bm{S}_i/2m_i$, we expect that the mixing effect characterized by Eq.~(\ref{mat_ele_1}) also becomes smaller.

Results for the transverse vector bottomonia [$\Upsilon(1S)_{|1 \pm 1 \rangle}$, $\Upsilon(2S)_{|1 \pm 1 \rangle}$ and $\Upsilon(3S)_{|1 \pm 1 \rangle}$] are shown in Figs.~\ref{Results_bottomT_m} and \ref{Results_bottomT_r}.
We find that the mass shifts and the wave function deformation of bottomonia are less sensitive to a magnetic field than those of charmonia.
For instance, the mass shifts of $\Upsilon(1S)$, $\Upsilon(2S)$ and $\Upsilon(3S)$ at $eB=0.3\mathrm{GeV}^2$ are $+0.4$, $+2.3$ and $+5.0\mathrm{MeV}$.
These smaller mass shifts could be experimentally irrelevant in heavy-ion collisions.
Also, the shapes of the wave functions are almost spherical even at $eB=1.0 \ \mathrm{GeV}^2$.
We find that first the mass of the ground state increases loosely and finally changes to the linear increase by the quark Landau levels.
Such linear behavior can also be shown for the excited state in weaker magnetic fields.
For the RMS radii shown in Fig.~\ref{Results_bottomT_r}, while the radii of the ground state in both the $\rho$- and $z$-directions decease, for the excited states $2S$ and $3S$, we find the shrinkage in the $\rho$-plane and the expansion in the $z$-direction.
In particular, the radii of $\Upsilon(2S)$ and $\Upsilon(3S)$ in the $\rho$-plane approach to the same value as that of $\Upsilon(1S)$ because the excitation in the $\rho$-plane is removed by a magnetic field and only that in the $z$-direction survives.
From the result, in the strong field limit, we expect that the $\rho$-radii of all the states reach to the same saturation value for a magnetic field.
On the other hand, the excitations in the $z$-direction behave as a quasi-one-dimensional object with a confinement potential.  

The results of the pseudoscalar [$\eta_b(1S)$, $\eta_b(2S)$ and $\eta_b(3S)$] and longitudinal vector [$\Upsilon(1S)_{|1 0 \rangle} $, $\Upsilon(2S)_{|1 0 \rangle}$ and $\Upsilon(3S)_{|1 0 \rangle}$] bottomonia are shown in Figs.~\ref{Results_bottomL_m} and \ref{Results_bottomL_r}.
For the {\it first} state starting from $\eta_b(1S)$ in vacuum, the modification in a magnetic field is less sensitive than that of the corresponding state in charmonium.
For instance, the mass shift and the mixing rate at $eB=0.3 \, \mathrm{GeV}^2$ are $-5.9 \, \mathrm{MeV}$ and $8\%$, respectively.
Such behavior is quantitatively consistent with the previous studies based on the same Hamiltonian in Refs.~\cite{Alford:2013jva,Bonati:2015dka}.
The mass of the {\it second} state starting from $\Upsilon(1S)$ increases gradually by the level repulsion and approaches to the third state at about $eB=5.5 \, \mathrm{GeV}^2$.
After the crossing, the wave function like the ground state whose $\sqrt{\langle z^2 \rangle} \simeq 0.2 \, \mathrm{fm}$ converts to the $2S$-excitation in the $z$-direction, of which $\sqrt{\langle z^2 \rangle} = 0.7 \, \mathrm{fm}$.
Also, the mixing rate having been $45\%$ changes to almost $50\%$.
The mass of the {\it third} state does not change almost up to $eB=5.5 \, \mathrm{GeV}^2$.
In this region, the wave function in the $z$-direction is gradually expanded.
After that, it approaches the second state like $\Upsilon (1S)$ and the third state like $\eta_c(3S)$ at $eB=8.5 \, \mathrm{GeV}^2$.
These crossing magnetic fields seem to be quite larger than the case of the charmonia, where the corresponding fields are $eB=1.0$-$1.1 \, \mathrm{GeV}^2$ and $eB=1.8$-$1.9 \, \mathrm{GeV}^2$.
The upward mass shifts of the {\it fourth} state starting from $\Upsilon(2S)$ in weak magnetic fields are larger than those of the first state.
The state approaches the fifth one like $\eta_c(3S)$, and, after the crossing at $eB=3.0 \, \mathrm{GeV}^2$, the mass is almost unchanged.
Moreover, after the crossing to the third state, the mass increases and will approach to the fifth state again.
The behavior of the {\it fifth} ({\it sixth}) state starting from $\eta_b(3S)$ [$\Upsilon(3S)$] is similar to the third (fourth) state.
Then we stress that the $B\bar{B}$ threshold in vacuum is located at about $10.56 \, \mathrm{GeV}$, so that the modifications of bottomonia up to $\Upsilon(3S)$ in the weak magnetic field can be estimated without the threshold effect.
In a stronger magnetic field, we need to take into account the competition with threshold effects related to the magnetic field dependences of $B$ mesons as shown in the next section.

\subsection{Heavy-light mesons}
In this section, we investigate the neutral heavy-light mesons: $D$, $B$ and $B_s$ mesons.
As with quarkonia, heavy-light meson systems exhibit the mixing between $| 00 \rangle$ and $| 10 \rangle$ components \cite{Machado:2013rta,Gubler:2015qok}.
In addition, these mesons consist of one heavy quark and one light antiquark and have a nonzero magnetic moment, which induces the Zeeman splitting between the transverse components of the spin triplet ($| 1 +1 \rangle$ and $| 1 -1 \rangle$) in a magnetic field.
As a result, the three states of the spin triplet, which degenerate in vacuum, are completely separated in a magnetic field. 

It is important that these mesons include one (constituent) light quark so that one of the origins of such a constituent quark mass should come from the chiral symmetry breaking.
Since a magnetic field enhances the chiral symmetry breaking by the magnetic catalysis, we may actually take into account the effect in our analyses for mesons with a light quark.
In this section, we show meson systems from the model {\it without} the magnetic catalysis.
Analyses with the magnetic catalysis will be discussed in the next section.

\begin{figure}[t!]
    \begin{minipage}[h]{1.0\columnwidth}
        \centering
        \includegraphics[clip, width=1.0\columnwidth]{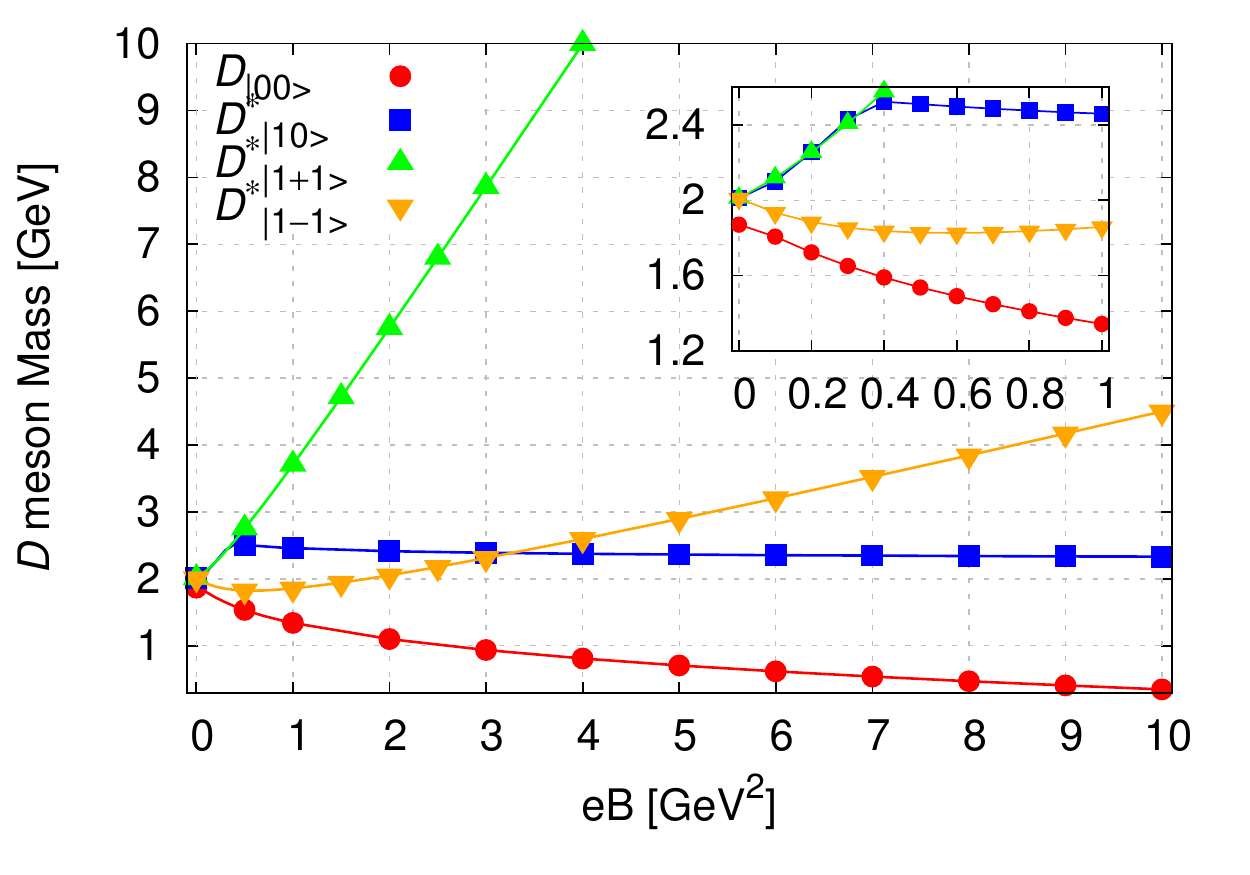}
    \end{minipage}
    \begin{minipage}[h]{1.0\columnwidth}
        \centering
        \includegraphics[clip, width=1.0\columnwidth]{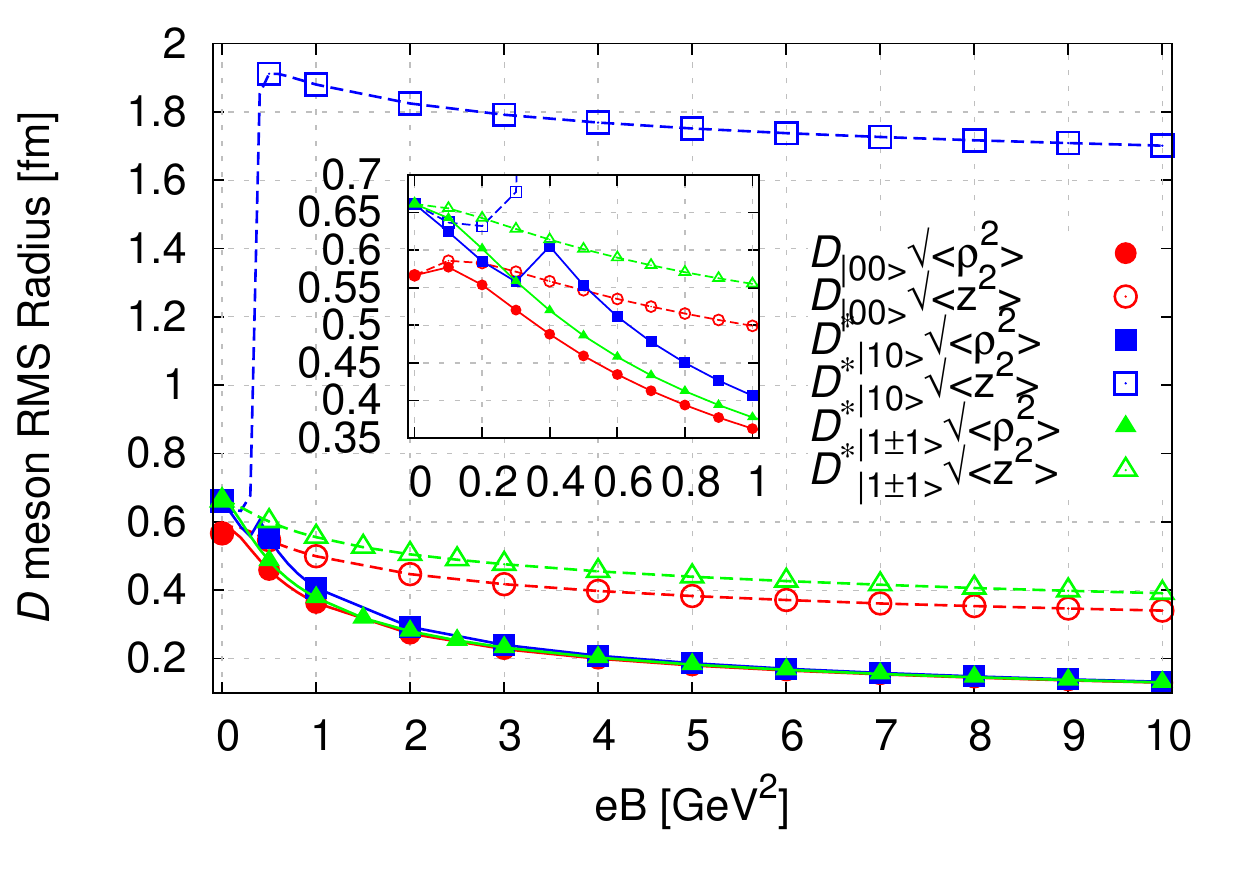}
    \end{minipage}
    \begin{minipage}[h]{1.0\columnwidth}
        \centering
        \includegraphics[clip, width=1.0\columnwidth]{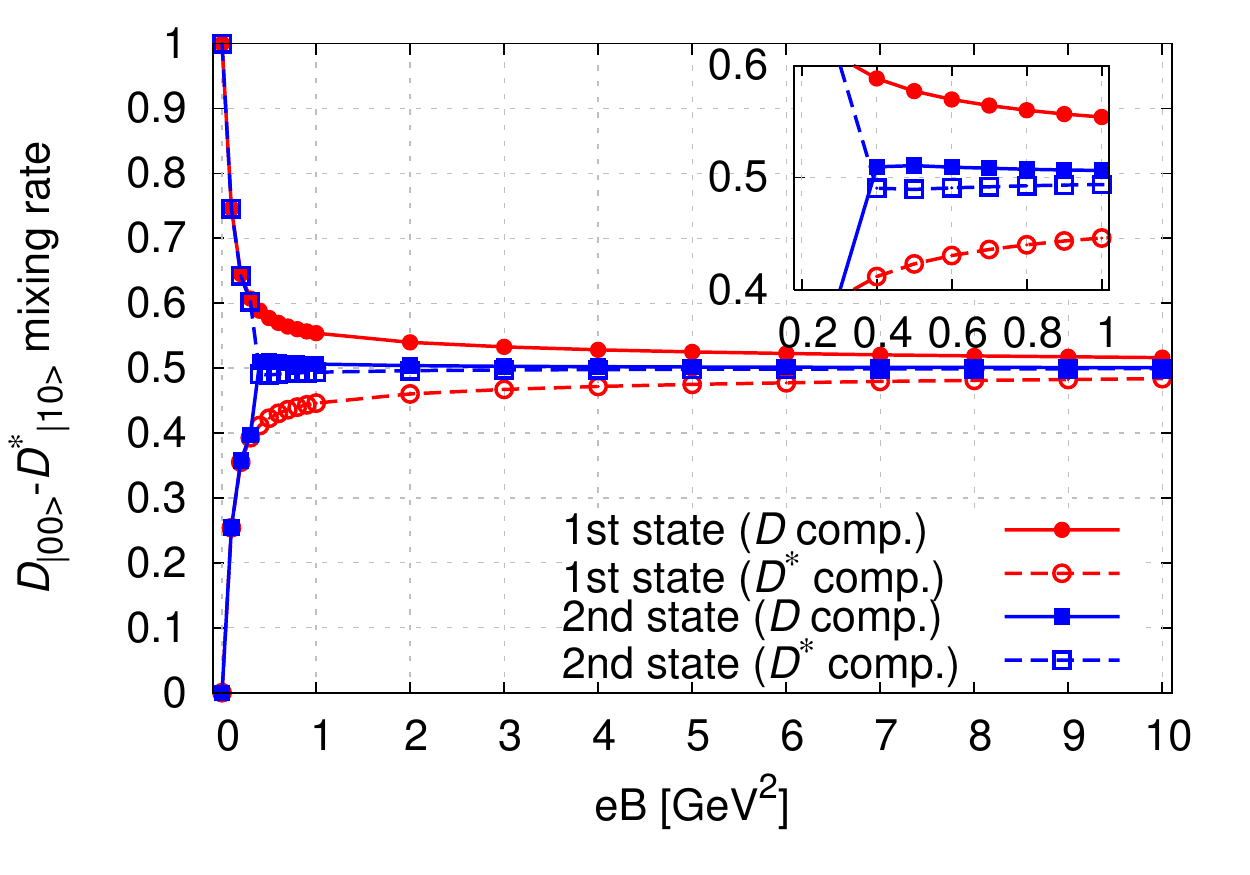}
    \end{minipage}
    \caption{Masses, RMS radii and mixing rates of neutral $D$ and $D^\ast$ mesons in a magnetic field.}
    \label{Results_D}
\end{figure}

\begin{figure}[t!]
    \begin{minipage}[h]{1.0\columnwidth}
        \centering
        \includegraphics[clip, width=1.0\columnwidth]{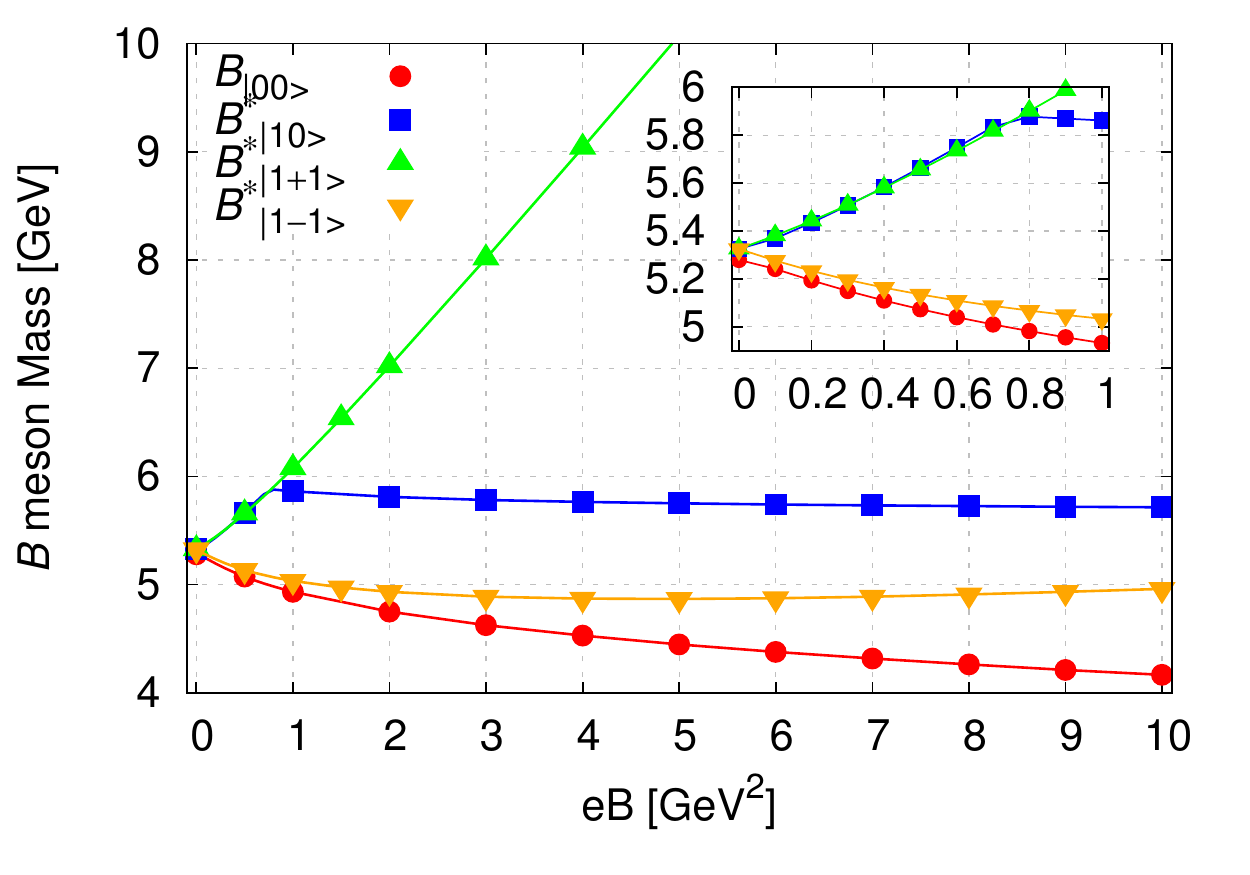}
    \end{minipage}
    \begin{minipage}[h]{1.0\columnwidth}
        \centering
        \includegraphics[clip, width=1.0\columnwidth]{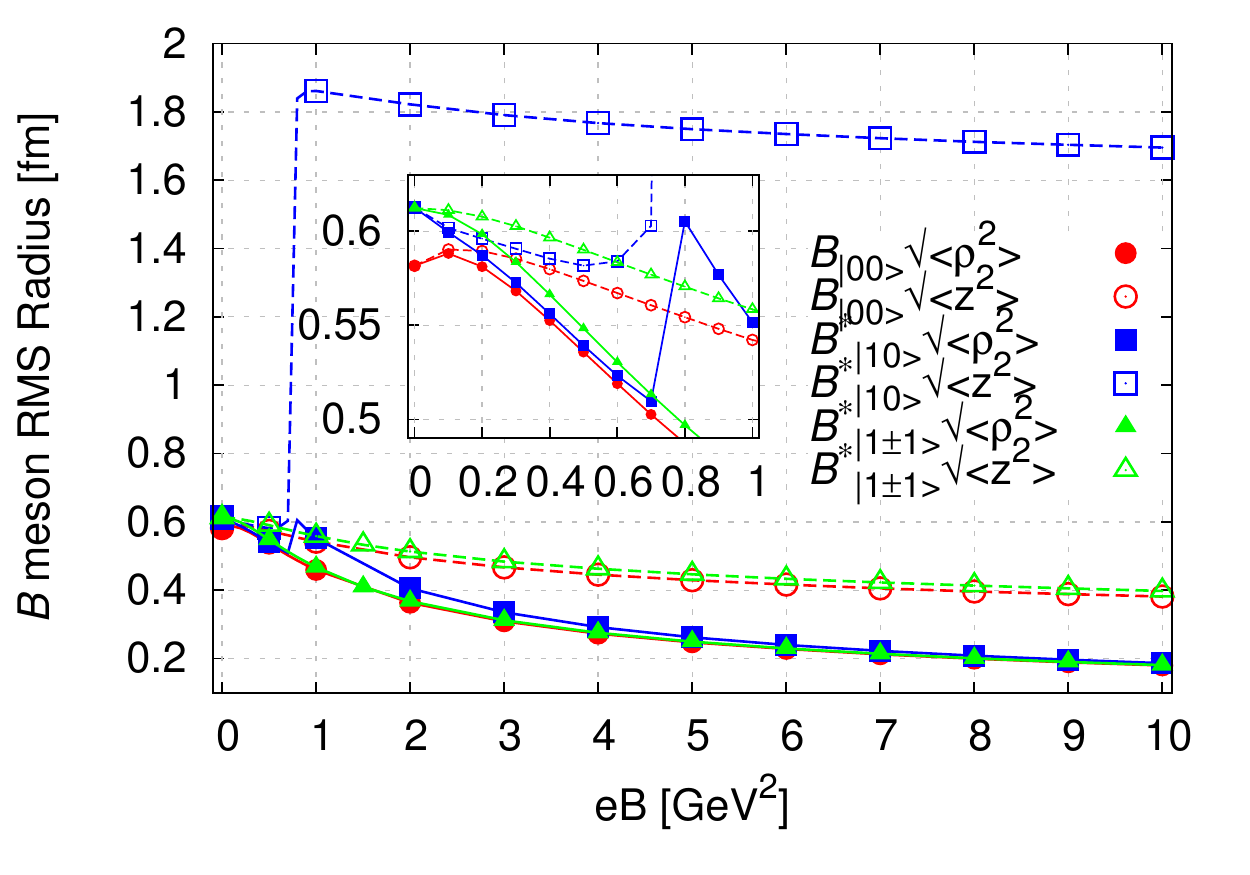}
    \end{minipage}
    \begin{minipage}[h]{1.0\columnwidth}
        \centering
        \includegraphics[clip, width=1.0\columnwidth]{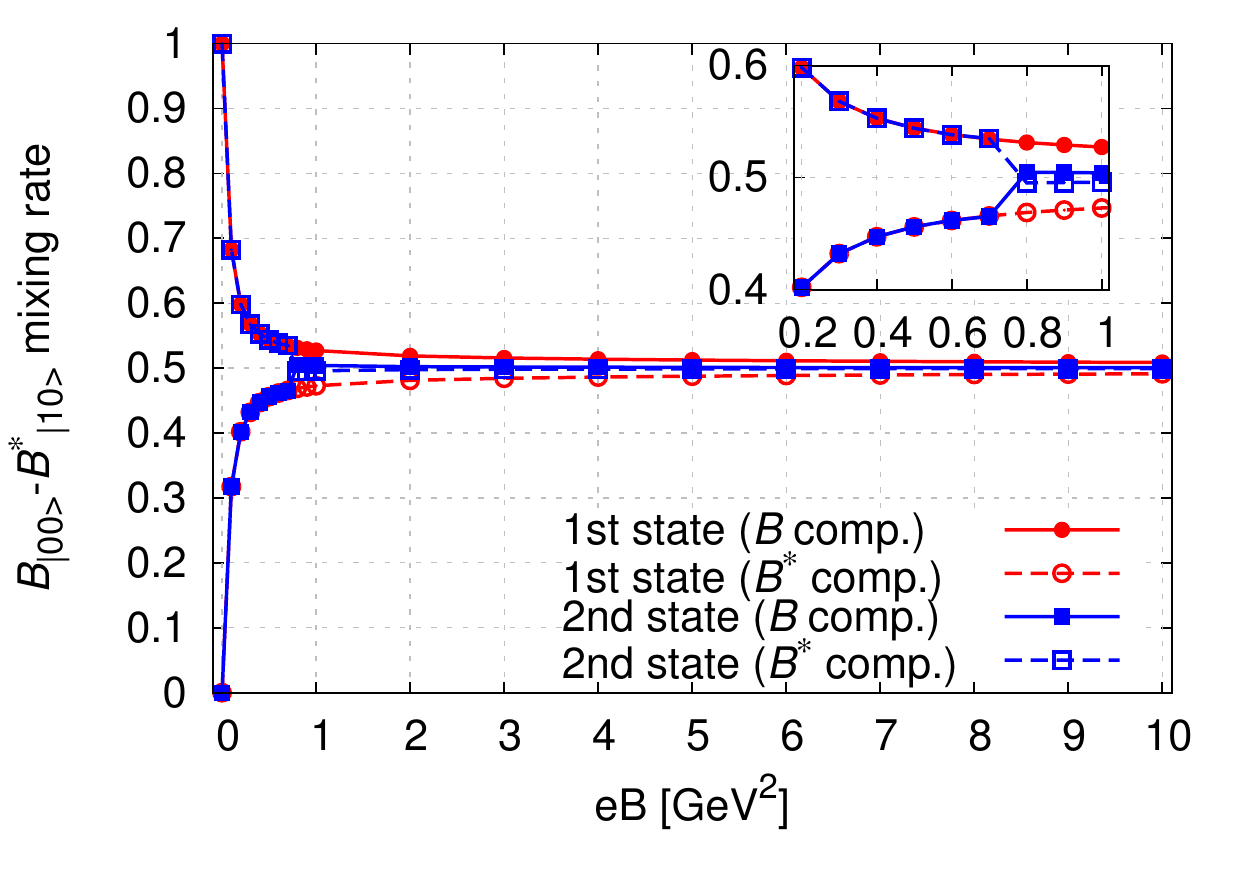}
    \end{minipage}
    \caption{Masses, RMS radii and mixing rates of neutral $B$ and $B^\ast$ mesons in a magnetic field.}
    \label{Results_B}
\end{figure}

\begin{figure}[t!]
    \begin{minipage}[h]{1.0\columnwidth}
        \centering
        \includegraphics[clip, width=1.0\columnwidth]{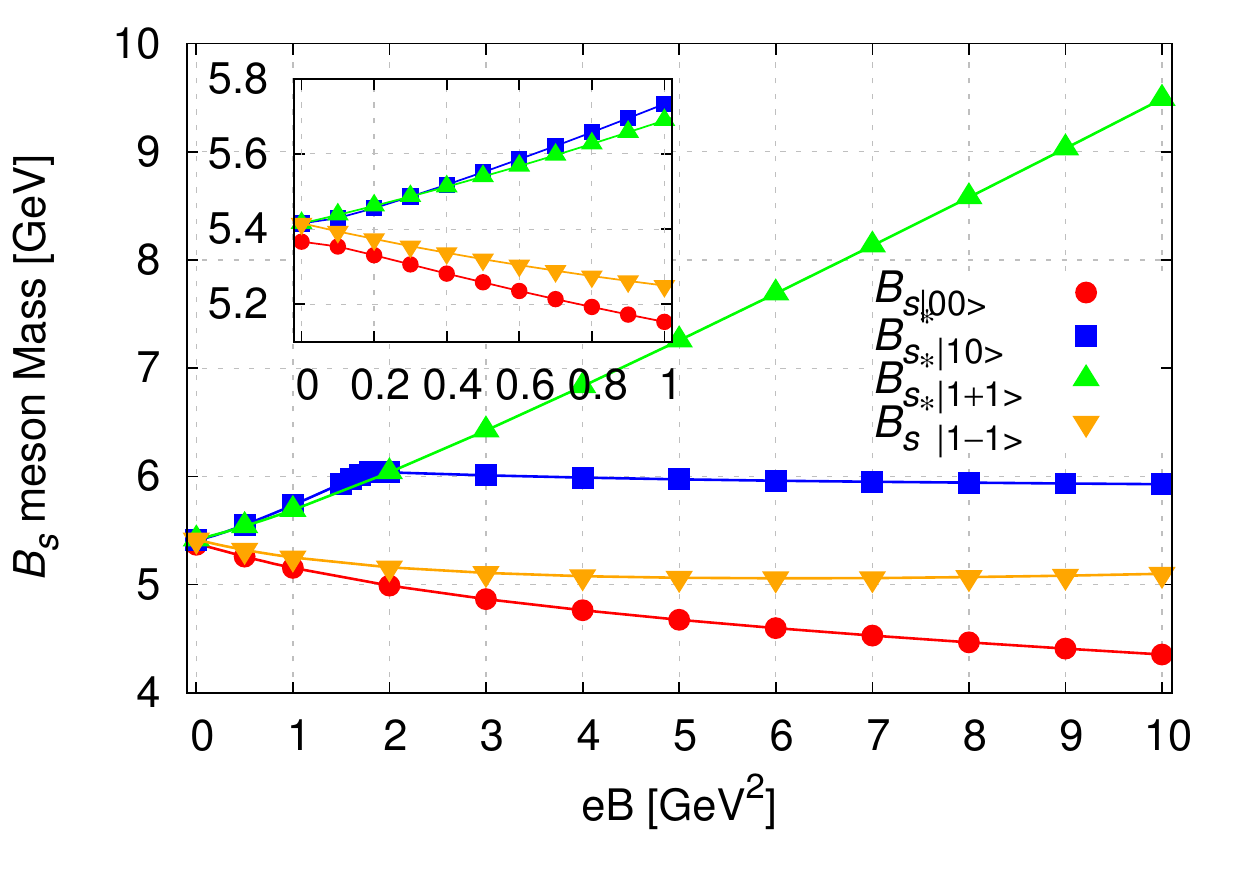}
    \end{minipage}
    \begin{minipage}[h]{1.0\columnwidth}
        \centering
        \includegraphics[clip, width=1.0\columnwidth]{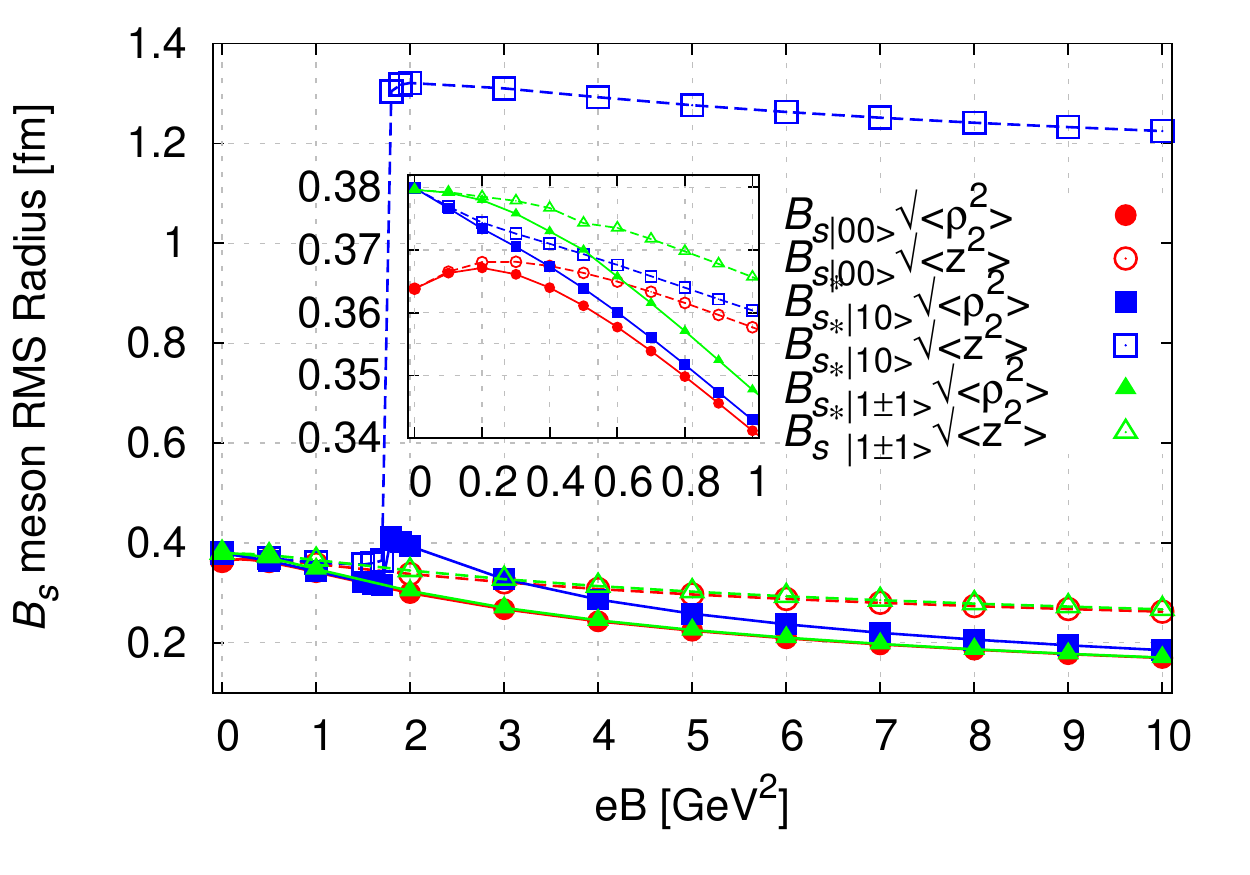}
    \end{minipage}
    \begin{minipage}[h]{1.0\columnwidth}
        \centering
        \includegraphics[clip, width=1.0\columnwidth]{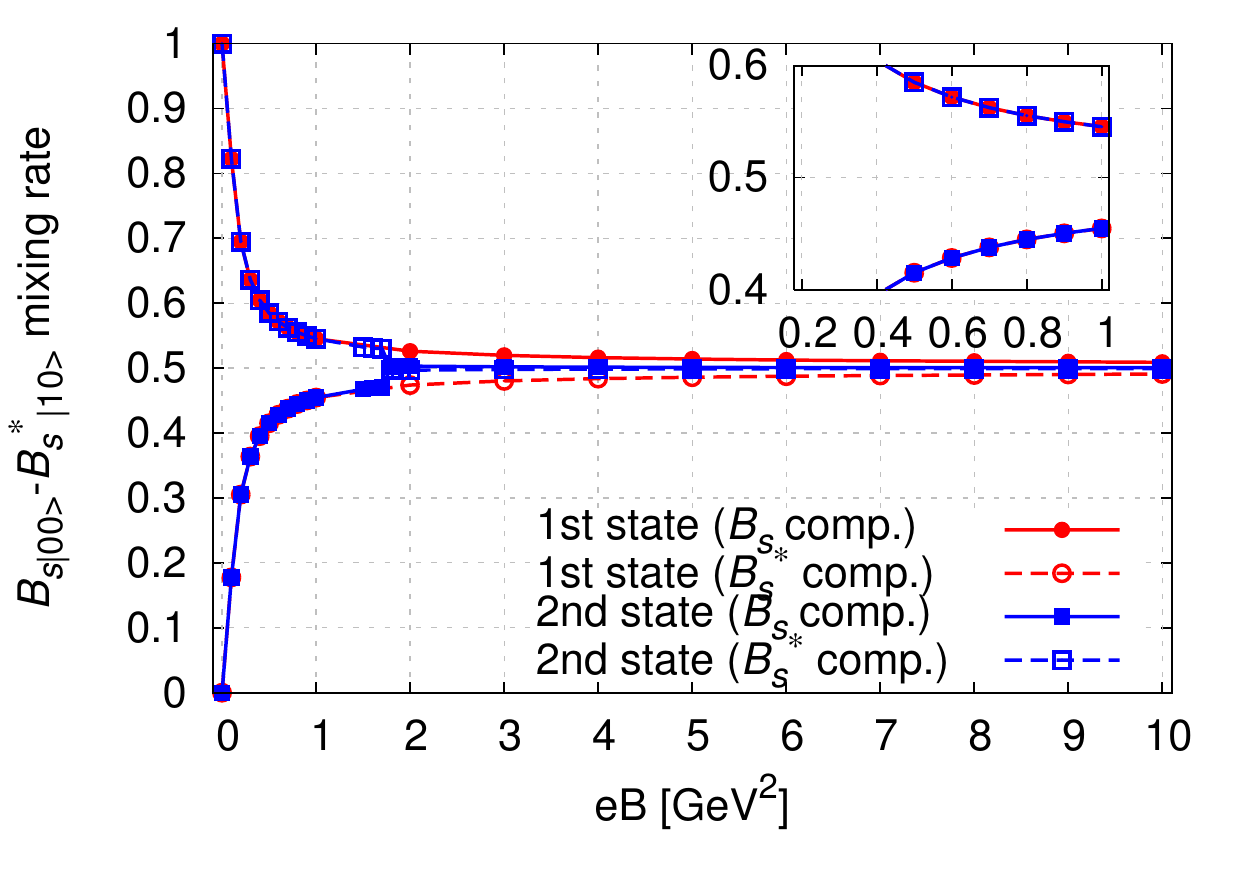}
    \end{minipage}
    \caption{Masses, RMS radii and mixing rates of neutral $B_s$ and $B_s^\ast$ mesons in a magnetic field.}
    \label{Results_Bs}
\end{figure}

Results for $D$, $B$ and $B_s$ mesons are shown in Figs.~\ref{Results_D}, \ref{Results_B}, and \ref{Results_Bs}, respectively.
In a weak magnetic field, the transverse vector heavy-light mesons ($D^\ast_{|1 \pm1 \rangle}$, $B^\ast_{|1 \pm1 \rangle}$, and $B^\ast_{s |1 \pm1 \rangle}$) are affected by the Zeeman splitting induced by the matrix element Eq.~(\ref{mat_ele_2}) and $s_z=\pm1$ states which degenerate in vacuum are split.
Namely, the masses of upper states (symbolized as green points) increase and those of lower states (symbolized as orange points) decrease. 
In stronger magnetic fields, the contribution from quark Landau levels becomes larger and comparable with that from the Zeeman splitting.
The masses of the upper states continue to increase.
On the other hand, in the masses of the lower states, the quark Landau levels balance with the Zeeman splitting and the masses begin to increase at a ``critical'' magnetic field, at $eB=0.6$, $5.0$, $ 6.0 \, \mathrm{GeV}^2$ for $D^\ast$, $B^\ast$, and $B^\ast$, respectively.
$s_z=\pm1$ states have the same RMS radii because the operator $-\bm{\mu}_i \cdot \bm{B}$ only contributes to spin wave functions and does not affect to spatial wave functions.
From the middle figures in Figs.~\ref{Results_D}, \ref{Results_B}, and \ref{Results_Bs}, we find that the RMS radii of the ground states of transverse vector heavy-light mesons continue to decrease.

Pseudoscalar ($D$, $B$, $B_s$) and longitudinal vector ($D^\ast_{|1 0 \rangle}$, $B^\ast_{|1 0 \rangle}$, $B^\ast_{s|1 0 \rangle}$) mesons under magnetic field are mixed with each other, which is similar to quarkonia.
Also, we can see the similar spectroscopic structures including the level crossing.
Furthermore, the mass shifts of the upper states of the transverse vector mesons (green points) and the longitudinal vector mesons (blue points) almost agree with each other in weak fields.
This behavior can be simply understood by considering the form of
the matrix elements of Eqs.~(\ref{mat_ele_1}) and (\ref{mat_ele_2}) near the heavy quark limit ($m_2 \to \infty$).
In the limit, the diagonal component by the $-\mu\cdot B$ term in transverse vector mesons agrees with
the off-diagonal components in the longitudinal vector mesons.
Therefore, as shown by the simplified two-level model, the mass increase by the mixing is the same as
that by the Zeeman splitting.

For all the heavy meson channels, the evaluated ``level crossing magnetic fields'' are summarized in Table \ref{crossing}. 
Here, it is interesting that we compare $D$ ($B$ and $B_s$) mesons with charmonia (bottomonia), where the quark electric charges are the same and the only difference is one light quark mass.
The crossing field in the $D$ ($B$ and $B_s$) meson channel is about $3$ ($8$ and $3$) times smaller than in the charmonium (bottomonium) channel.
These ratios are almost consistent with the strength of the matrix element Eq.~(\ref{mat_ele_1}) which is proportional to the first order of $|eB|$ [$\propto |eB|(1/m_1 +1/m_2)$].  
\begin{table*}[t!]
   \begin{center}
   \caption{List of level crossing magnetic fields for pseudoscalar-vector mixing states in heavy mesons in a magnetic field, which is evaluated by the constituent quark model, Eq.~(\ref{H_rel2}), and CGEM.
For heavy-light mesons, contributions from the magnetic catalysis are {\it not} taken into account.}
   \label{crossing}
  \begin{tabular}{l|cccccc}
\hline \hline
Charmonium                   & $J/\psi$-$\eta_c(2S)$ & $\eta_c(2S)$-$\psi(2S)$ & $\psi(2S)$-$\eta_c(3S)$ &  & &\\
\hline
Crossing $B$-field [GeV$^2$] & $1.0$-$1.1$             & $1.8$-$1.9$               & $0.6$-$0.7$, $2.4$-$2.5$    &  & \\
\hline \hline
Bottomonium                  & $\Upsilon(1S)$-$\eta_b(2S)$ & $\eta_b(2S)$-$\Upsilon(2S)$ & $\Upsilon(2S)$-$\eta_b(3S)$ & $\eta_b(3S)$-$\Upsilon(3S)$ & $\Upsilon(3S)$-$\eta_b(4S)$  \\
\hline
Crossing $B$-field [GeV$^2$] & $5.5$                 & $8.5$                   & $3.0$                   & $5.0$ & $2.5$, $6.5$ \\
\hline \hline
Heavy-light meson            & $D^\ast(1S)$-$D(2S)$  & $B^\ast(1S)$-$B(2S)$    & $B_s^\ast(1S)$-$B_s(2S)$& &\\
\hline
Crossing $B$-field [GeV$^2$] & $0.3$-$0.4$           & $0.7$-$0.8$             & $1.7$-$1.8$             & & \\
\hline \hline
   \end{tabular}
   \end{center}
\end{table*}
This can be checked by a replacement of one quark mass, $m_c \to m_q$ ($m_b \to m_q$ or $m_s$).
Furthermore, RMS radii and mixing rates of $D$ ($B$) mesons are more sensitive than the corresponding charmonia (bottomonia).
As with the RMS radii of quarkonia, we also found the increases of radii of the first states in smaller magnetic fields.  
For the second states, we do not see such behavior.
We obtained the mixing rates of $39\%$, $43\%$ and $36\%$ at $eB=0.3 \, \mathrm{GeV}^2$ for $D$, $B$, and $B_s$ mesons, respectively.

\subsection{Magnetic catalysis in heavy-light mesons} \label{SubSec_MC}

As shown in Ref.~\cite{Park:2016xrw}, in the constituent quark model, the chiral symmetry restoration (or the reduction of $\langle \bar{q}q \rangle$) in the nuclear medium corresponds to smaller values of the constituent quark mass, which leads to an {\it increase} of the mass of a heavy-light meson.
Such a picture for $D$ meson masses at finite density can be understood \cite{Park:2016xrw} by a balance between a decrease of the constituent quark mass, $m_q$, and an increase of the kinetic and potential energies, $E_K+E_V$, which is consistent with the results from QCD sum rules \cite{Hilger:2008jg,Suzuki:2015est} and some phenomenological approaches \cite{Blaschke:2011yv,Sasaki:2014asa,Suenaga:2014sga}.
In contrast, in a magnetic field, the chiral symmetry breaking is enhanced by the magnetic catalysis and the constituent light quark mass should increase.
Therefore, we may naively expect that the mass of a heavy-light meson {\it decreases} with increasing magnetic field.
In this section, we investigate the light quark mass dependence of the analyses of the last section.

To introduce the $B$-field dependence of a constituent quark mass, we adopt the chiral condensate estimated from the two-flavor NJL model \cite{Boomsma:2009yk} as shown in Appendix \ref{App_NJL}, where we input the differences from the value in vacuum, $\langle \bar{q} q \rangle_{eB} - \langle \bar{q} q \rangle_{\mathrm{vac}}$. Here, the values of $\langle \bar{u} u \rangle$ and $\langle \bar{d} d \rangle$ are distinguished from each other and the instanton interaction is neglected.
Therefore, for analyses of neutral $D$ mesons composed of charm and up quarks, we use the values of $\langle \bar{u} u \rangle$.

\begin{figure}[t!]
    \begin{minipage}[h]{1.0\columnwidth}
        \centering
        \includegraphics[clip, width=1.0\columnwidth]{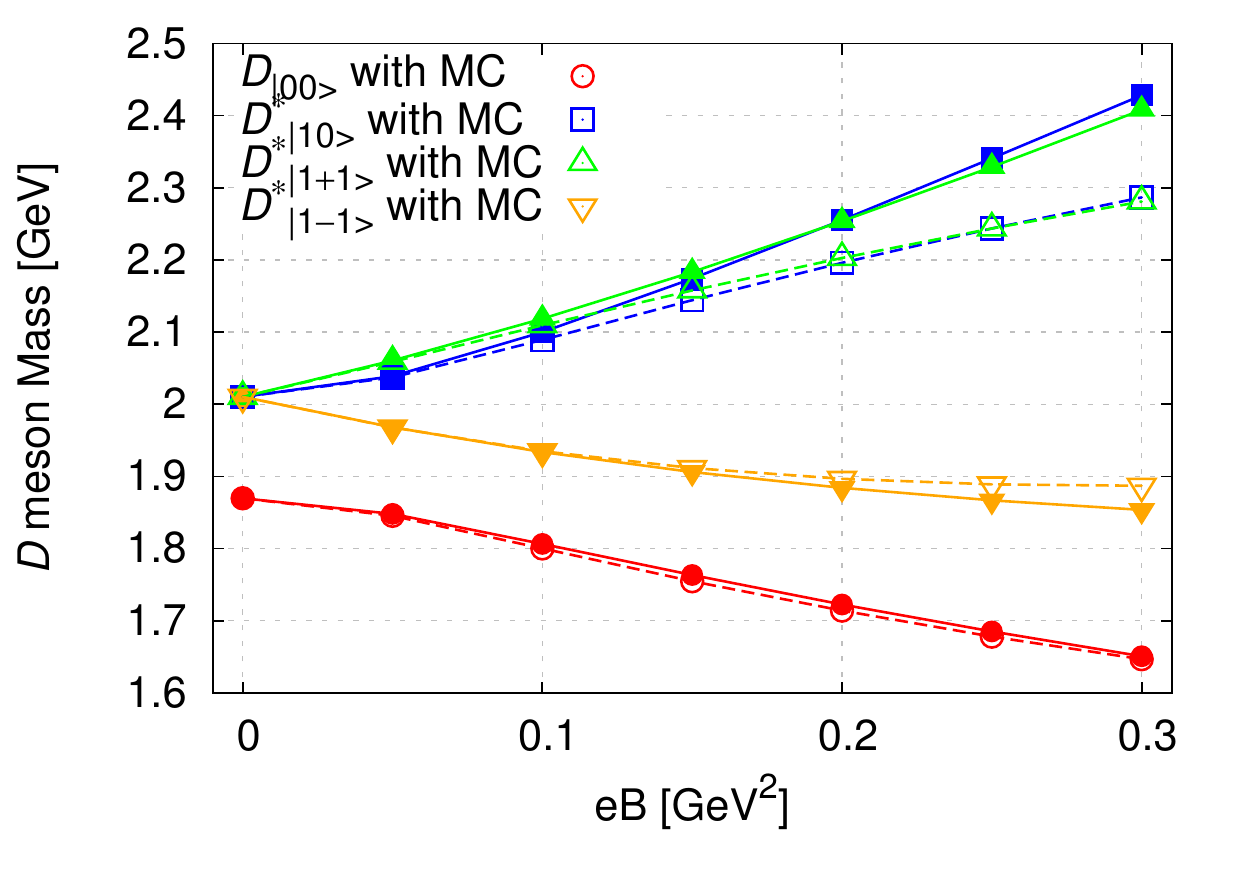}
    \end{minipage}
    \caption{Light quark mass dependences of masses of $D$ and $D^\ast$ mesons in a magnetic field up to $eB=0.3 \rm{GeV}^2$.
Dashed (solid) lines correspond to mass with (without) the magnetic catalysis (MC).
The solid lines are the same as the lines in Fig.~\ref{Results_D}.}
    \label{Results_D_MC}
\end{figure}

The $D$-meson masses with the $B$-field dependence of the light quark mass are shown in Fig.~\ref{Results_D_MC}. 
In this figure, the dashed lines correspond to the masses taking into account the magnetic catalysis.
The light quark mass can contribute to the two components: (i) the kinetic and potential energies $E_K+E_V$, and (ii) the magnetic-moment given by $\bm{\mu}_q \equiv g q_q \bm{S}_q/2m_q$.
As explained above, for (i), $E_K+E_V$ in a heavy-light meson decreases with increasing $m_q$.
On the other hand, for (ii), the magnetic moment of the light quark becomes smaller as $m_q$ increases, so that the strength of the spin mixing (or the Zeeman splitting) is weakened. [See also the form of the matrix elements, Eqs.~(\ref{mat_ele_1}) and (\ref{mat_ele_2})].
As a result, we find that the pseudoscalar $D$-meson mass is almost unchanged because of the cancellation between the mass decrease by (i) and the mass increase from the weaker level repulsion based on (ii).
The effect of (i) can also be applied to $D^\ast$ mesons.
The mass of the $| 10 \rangle$ component of $D^\ast$ decreases by the magnetic catalysis because of both the mass decreases by (i) and the smaller level repulsion.
Similarly, for the $| 1 \pm1 \rangle$ components of $D^\ast$, we can find the mass shifts which can be understood by the mass decrease by (i) and the reduction of the Zeeman splitting.

We note that the results in this section may theoretically be important in the sense that, in some previous works, a hadron effective model for $D$ mesons was used \cite{Gubler:2015qok} and such a model does not include contributions from the magnetic catalysis.
If we would like to modify this model to include contributions from the magnetic catalysis, our current results imply necessities of quantitative corrections for $D^\ast$ mesons.
Furthermore, the previous analyses by QCD sum rules in Refs.~\cite{Machado:2013yaa, Gubler:2015qok} have taken into account the magnetic catalysis of the $\langle \bar{q} q \rangle$ condensate.
We additionally mention that hadron masses from QCD sum rules in a magnetic field can be contributed from not only $\langle \bar{q} q \rangle$ but also other condensates (e.g. $\langle \bar{q} \sigma_{\mu \nu} q \rangle$ as included in Ref.~\cite{Gubler:2015qok}) more or less, so the relation of such condensates to the quark model would be one of the future issues.  

\section{Conclusion and outlook} \label{Sec_Conclusion and outlook}
In this paper, we investigated properties of heavy meson (charmonia, bottomonia, neutral $D$, $B$, and $B_s$ mesons) spectra in a magnetic field.
We developed the method to solve the two-body Schr\"{o}dinger equation for a quark model in the cylindrical coordinate.
By using this approach, our results can include full contributions from higher excited states.
As a result, we found mass shifts and deformation of wave functions in heavy meson systems in a magnetic field.

For transverse vector quarkonia [$J/\psi_{|1 \pm1 \rangle}$, $\psi(2S)_{|1 \pm1 \rangle}$, $\Upsilon(1S)_{|1 \pm1 \rangle}$, $\Upsilon(2S)_{|1 \pm1 \rangle}$, and $\Upsilon(3S)_{|1 \pm1 \rangle}$], we found that there are mass shifts  although these hadrons have no charge.
On the other hand, for pseudoscalar and longitudinal vector charmonia [$\eta_c(1S)$, $J/\psi_{|1 0 \rangle}$, $\eta_c(2S)$, and $\psi(2S)_{|1 0 \rangle}$] and bottomonia [$\eta_b(1S)$, $\Upsilon_{|1 0 \rangle}(1S)$, $\eta_b(2S)$, $\Upsilon(2S)_{|1 0 \rangle}$, $\eta_b(3S)$, and $\Upsilon(3S)_{|1 0 \rangle}$], the spin partners (spin-0 and spin-1) mix with each other, so that we obtained the level structures with the level repulsion.

For the transverse vector heavy-light mesons ($D^\ast_{|1 \pm1 \rangle}$, $B^\ast_{|1 \pm1 \rangle}$, and $B^\ast_{s |1 \pm1 \rangle}$), we discussed mass shifts by Zeeman splitting.
For the pseudoscalar and longitudinal vector heavy-light mesons ($D$, $D^\ast_{|1 0 \rangle}$, $B$, $B^\ast_{|1 0 \rangle}$, $B_s$, $B^\ast_{s|1 0 \rangle}$), we examined the level structures induced by the mixing.
Moreover, for $D$ meson masses, the modification of the constituent quark mass based on the magnetic catalysis is discussed.
Here, we stress that heavy-light mesons under magnetic field can be useful for a probe of the magnetic catalysis.
In association with the magnetic catalysis, we can also expect hadron modifications from the gluonic magnetic catalysis \cite{Bali:2013esa,Ozaki:2013sfa} because properties of quarkonia are partially attributed to the gluon condensate instead of the chiral condensate.

It is worthwhile to note possible corrections in our approach: (i) {\it $B$-dependences of the model parameters}, (ii) {\it relativistic corrections}, and (iii) {\it threshold and continuum effects}.
(i) As stated in Sec. \ref{subsec_parameter}, in this work, only the $B$-dependences of the constituent light quark mass $m_q$ for $D$-meson channels were investigated.
The anisotropic linear and Coulomb potentials were evaluated from Wilson loops in a recent lattice QCD simulation \cite{Bonati:2014ksa}.
The applications of such potentials to the quark model are interesting, as performed in Ref.~\cite{Bonati:2015dka}.
For example, because, in Ref.~\cite{Bonati:2014ksa}, the linear potential becomes larger (smaller) in the $\rho$-direction ($z$-direction), we can naively expect the squeezing of wave functions in the $\rho$-direction to become more drastic than the current results.
In addition, the flavor dependence of the anisotropic potential and the $B$-dependences of the spin-spin interaction will also be discussed in future works.
(ii) Our approach leads to the physics in a nonrelativistic limit and its validity should be discussed.
The applicable region of the approach could be naively estimated from a ratio of the constituent quark mass $m$ and the strength of magnetic field (e.g. $eB/m^2 \ll 1$).
In strong fields beyond the applicable region, one has to consider the connection with the relativistic approach.
(iii) Some of the excited states investigated in this work are located near the continuum threshold, so that we need the treatments of threshold (or continuum) effects in a magnetic field.
Some possibilities of charmonium continuum in a magnetic field were discussed in Ref.~\cite{Cho:2014loa}.

As more realistic situations in heavy-ion collisions, we can investigate hadron properties at finite temperature (and/or density) in a magnetic field.
In the constituent quark model, a temperature (density) dependence can be introduced as changes of model parameters and functional forms.
A representative example is the potential model approach for $J/\psi$ suppression near the critical temperature \cite{Matsui:1986dk}.
The approach can be simply extended to thermal quarkonia in a magnetic field.
In addition, as discussed in this work, $D$ meson mass shifts are expected to be a probe of the magnetic catalysis so that studying the temperature (or density) dependence of the $\langle \bar{q} q \rangle$ condensate would be interesting \cite{Klimenko:1991he,Agasian:2001ym,Ebert:2003yk,Inagaki:2003yi,Agasian:2008tb,Fraga:2008qn,Menezes:2008qt,Boomsma:2009yk,Menezes:2009uc,Fukushima:2010fe,Mizher:2010zb,Gatto:2010qs,Fayazbakhsh:2010bh,Gatto:2010pt,Preis:2010cq,Chatterjee:2011ry,Frasca:2011zn,Rabhi:2011mj,Kashiwa:2011js,Andersen:2011ip,Skokov:2011ib,Andersen:2012dz,Fukushima:2012xw,Andersen:2012bq,Fayazbakhsh:2012vr,Fayazbakhsh:2013cha,Ballon-Bayona:2013cta,Andersen:2013swa,Fraga:2013ova,Tawfik:2014hwa,Andersen:2014xxa,Tawfik:2015tga,Mamo:2015dea,Hattori:2015aki} and would also be related to the experimental explorations of the inverse magnetic catalysis scenarios (e.g. see Refs. \cite{Fukushima:2012kc,Kojo:2012js,Bruckmann:2013oba,Chao:2013qpa,Ferreira:2013tba,Kamikado:2013pya,Farias:2014eca,Ferreira:2014kpa,Yu:2014sla,Ayala:2014iba,Ayala:2014gwa,Ferrer:2014qka,Andersen:2014oaa,Yu:2014xoa,Mueller:2015fka}) observed by lattice QCD simulations \cite{Bali:2011qj,Bali:2012zg,Bali:2013esa,Bruckmann:2013oba,Bornyakov:2013eya,Bali:2014kia}.

Our results in the weak magnetic fields are useful to predict the observables in a magnetic field induced by heavy-ion collisions, such as the $\eta_c + \gamma \to J/\psi$ transition rate \cite{Yang:2011cz}, production cross section of quarkonia \cite{Machado:2013rta}, anisotropic production \cite{Guo:2015nsa} and quarkonium melting \cite{Marasinghe:2011bt,Dudal:2014jfa}.
In particular, the dilepton spectra constituting decay modes such as $J/\psi, \psi^\prime, \Upsilon \to e^+ e^-$ can shift by a magnetic field and can exhibit a splitting between the transverse and longitudinal components \cite{Alford:2013jva,Filip:2013xza}.
Simultaneously, ``anomalous'' channels such as $\eta_c, \eta_b \to e^+ e^-$ can be opened due to the mixing with $J/\psi$ components, and these would be the signal of a magnetic field.

Recently, it was suggested that the isospin (QCD) Kondo effect in heavy-light mesons (heavy quarks) at finite density with a Fermi surface can occur and may modify transport properties of the systems \cite{Yasui:2013xr,Hattori:2015hka,Yasui:2016ngy,Yasui:2016svc} (see a part of Ref.~\cite{Hosaka:2016ypm} for a review).
Under the existence of a magnetic field, the magnetically induced Kondo effect can also be predicted \cite{Ozaki:2015sya}.
Properties of heavy quark systems with such effects are one of the more promising topics in heavy hadron physics.

\begin{acknowledgments}
The authors gratefully thank Makoto Oka, Sho Ozaki, Aaron Park and Shigehiro Yasui for fruitful discussions. K.S. was partially supported by a Grant-in-Aid for JSPS Fellows from the Japan Society for the Promotion of Science (JSPS) (Grant No. 26-8288). T.Y. acknowledges the Junior Research Associate scholarship
at RIKEN.
\end{acknowledgments}

\appendix
\begin{widetext}
\section{Matrix elements} \label{App_mat}
In this appendix, we summarize the matrix elements used in the cylindrical Gaussian expansion method.
We emphasize that to obtain the analytic forms of the matrix elements enables us to reduce the computational costs of the variational method.

\subsection{Norm matrix}
Each component $\Phi_n (\rho,z,\phi)$ for the expanded wave function is given in Eq.~(\ref{basis2}).
Then the corresponding norm matrix is as follows:
\begin{eqnarray}
N_{n n^\prime} &=& \langle \Phi_n | \Phi_{n^\prime} \rangle \nonumber\\
&=& \int_0^\infty d\rho \int_{-\infty}^\infty dz \int_0^{2\pi} d\phi \ \rho \Phi^*_n (\rho,z,\phi) \Phi_{n^\prime} (\rho,z,\phi) \nonumber\\
&=& N_n N_{n^\prime} \frac{\pi^{3/2}}{\sqrt{\gamma_n +\gamma_{n^\prime}}(\beta_n + \beta_{n^\prime})}.
\end{eqnarray}
By imposing the normalization $\langle \Phi_n | \Phi_{n} \rangle = 1$, we obtain the norm constant:
\begin{equation}
N_n =  \frac{2^{3/4} \gamma_n^{1/4} \sqrt{\beta_n}}{\pi^{3/4}}.
\end{equation}
Finally, we obtain 
\begin{equation}
N_{n n^\prime} = \frac{2^{3/2} (\gamma_n \gamma_{n^\prime})^{1/4} \sqrt{\beta_n \beta_{n^\prime}}}{\sqrt{\gamma_n +\gamma_{n^\prime}}(\beta_n + \beta_{n^\prime})}.
\end{equation}

\subsection{Kinetic term}
By using the Laplacian in the cylindrical coordinate,
\begin{equation}
\nabla^2 = \frac{\partial^2}{\partial \rho^2} + \frac{1}{\rho} \frac{\partial}{\partial \rho} +\frac{1}{\rho^2} \frac{\partial^2}{\partial \phi^2} + \frac{\partial^2}{\partial z^2},
\end{equation}
we obtain the matrix element for the kinetic term:
\begin{eqnarray}
T_{n n^\prime}^{B=0} &\equiv& \langle \Phi_n |  -\frac{1}{2\mu} \nabla^2 |\Phi_{n^\prime} \rangle \nonumber\\
&=&  -\frac{1}{2\mu} \left( N_n N_{n^\prime} \frac{\pi^{3/2}}{\sqrt{\gamma_n +\gamma_{n^\prime}}(\beta_n + \beta_{n^\prime})} \right) \left[ \frac{ - 2\beta_n \beta_{n^\prime} + 2\beta_{n^\prime}^2 }{\beta_n + \beta_{n^\prime}} -2 \beta_{n^\prime} + \frac{-2 \gamma_n \gamma_{n^\prime} }{\gamma_n +\gamma_{n^\prime}}  \right].
\end{eqnarray}

Also, the $B$-dependent kinetic term is
\begin{eqnarray}
T_{n n^\prime}^{B} &\equiv& \langle \Phi_n |  \frac{q^2 B^2}{8\mu} \rho^2 |\Phi_{n^\prime} \rangle \nonumber\\
&=&  \frac{q^2 B^2}{8\mu} N_n N_{n^\prime} \frac{\pi^{3/2}}{\sqrt{\gamma_n +\gamma_{n^\prime}} ( \beta_n + \beta_{n^\prime})^2 }.
\end{eqnarray}

\subsection{Potential terms}
The matrix elements for the potential terms (linear, Coulomb and spin-spin interactions) are given as
\begin{eqnarray}
V_{n n^\prime}^{\mathrm{linear}} &\equiv& \langle \Phi_n | \sigma \sqrt{\rho^2 + z^2} |\Phi_{n^\prime} \rangle \nonumber\\
&=& 2\pi \sigma N_n N_{n^\prime} \left \{
\begin{array}{ll}
\cfrac{1}{2\beta_{n n^\prime} \gamma_{n n^\prime}} \left( 1+ \cfrac{\gamma_{n n^\prime} \mathrm{arccosh}{\sqrt{\frac{\beta_{n n^\prime}}{\gamma_{n n^\prime}}}}}{\sqrt{\beta_{n n^\prime}(\beta_{n n^\prime}-\gamma_{n n^\prime})}} \right) & \mathrm{for} \ \beta_{n n^\prime}>\gamma_{n n^\prime} \\
\cfrac{1}{\gamma_{n n^\prime}^2} & \mathrm{for} \ \beta_{n n^\prime}=\gamma_{n n^\prime} \\
\cfrac{1}{2\beta_{n n^\prime} \gamma_{n n^\prime}} \left( 1+ \cfrac{\gamma_{n n^\prime} \arccos{\sqrt{\frac{\beta_{n n^\prime}}{\gamma_{n n^\prime}}}}}{\sqrt{\beta_{n n^\prime}(-\beta_{n n^\prime}+\gamma_{n n^\prime})}} \right) & \mathrm{for} \ \beta_{n n^\prime}<\gamma_{n n^\prime}
\end{array}, \right.
\end{eqnarray}
where $\beta_{n n^\prime} = \beta_{n} + \beta_{n^\prime}$ and $\gamma_{n n^\prime} = \gamma_{n} + \gamma_{n^\prime}$,

\begin{eqnarray}
V_{n n^\prime}^{\mathrm{Coulomb}} &\equiv& \langle \Phi_n | - \frac{A}{\sqrt{\rho^2 + z^2}} |\Phi_{n^\prime} \rangle \nonumber\\
&=& - A 2\pi N_n N_{n^\prime} \left \{
\begin{array}{ll}
\cfrac{\mathrm{arccosh} \left(-1 + \frac{2\beta_{n n^\prime}}{\gamma_{n n^\prime}} \right)}{2 \sqrt{\beta_{n n^\prime} (\beta_{n n^\prime} - \gamma_{n n^\prime})}} & \mathrm{for} \ \beta_{n n^\prime}>\gamma_{n n^\prime} \\
\cfrac{1}{\gamma_{n n^\prime}} & \mathrm{for} \ \beta_{n n^\prime}=\gamma_{n n^\prime} \\
\cfrac{\mathrm{arccos} \left(-1 + \frac{2\beta_{n n^\prime}}{\gamma_{n n^\prime}} \right)}{2 \sqrt{\beta_{n n^\prime} (-\beta_{n n^\prime} + \gamma_{n n^\prime})}} & \mathrm{for} \ \beta_{n n^\prime}<\gamma_{n n^\prime} <2 \beta_{n n^\prime} \\
\cfrac{\pi}{2\gamma_{n n^\prime}} &  \mathrm{for} \ 2 \beta_{n n^\prime}=\gamma_{n n^\prime} \\
\cfrac{\mathrm{arccos} \sqrt{\frac{\beta_{n n^\prime}}{\gamma_{n n^\prime}}}}{\sqrt{\beta_{n n^\prime} (-\beta_{n n^\prime} + \gamma_{n n^\prime})}} & \mathrm{for} \ 2 \beta_{n n^\prime}<\gamma_{n n^\prime}
\end{array}, \right.
\end{eqnarray}

\begin{eqnarray}
V_{n n^\prime}^{\mathrm{SS}} &\equiv& \langle \Phi_n | \alpha e^{- \Lambda (\rho^2 + z^2)} |\Phi_{n^\prime} \rangle \nonumber\\
&=& \alpha (\bm{S}_1 \cdot \bm{S}_2) N_n N_{n^\prime} \frac{\pi^{3/2}}{\sqrt{\gamma_n +\gamma_{n^\prime}+ \Lambda}(\beta_n + \beta_{n^\prime}+ \Lambda )}.
\end{eqnarray}

\subsection{Root-mean-square radii} \label{App_RMS}
In this work, we define the root-mean-square radii, $\langle \rho^2 \rangle $ and $\langle z^2 \rangle$, for the $\rho$- and $z$-directions, respectively.
\begin{eqnarray}
\sqrt{ \langle \rho^2 \rangle} &\equiv& \sqrt{ (3/2) \langle \Psi | \rho^2 |\Psi \rangle} \nonumber\\
&=& \sqrt{ \frac{3}{2} \sum_{n=1}^N \sum_{n^\prime=1}^N C_n C_{n^\prime} N_n N_{n^\prime} \frac{\pi^{3/2}}{\sqrt{\gamma_n +\gamma_{n^\prime}} ( \beta_n + \beta_{n^\prime})^2 }}, \\
\sqrt{ \langle z^2 \rangle} &\equiv& \sqrt{ 3 \langle \Psi | z^2 |\Psi \rangle } \nonumber\\
&=& \sqrt{ 3 \sum_{n=1}^N \sum_{n^\prime=1}^N C_n C_{n^\prime} N_n N_{n^\prime} \frac{\pi^{3/2}}{2 (\beta_n + \beta_{n^\prime}) (\gamma_n +\gamma_{n^\prime})^{3/2} }},
\end{eqnarray}
where the factors, $\sqrt{3/2}$ and $\sqrt{3}$, are introduced to compare the anisotropic radii with the spherical radius $\langle r^2 \rangle$ in vanishing magnetic field.
\end{widetext}

\section{Generalized eigenvalue problem for coupled-channel basis} \label{App_GEP}
Here we review the generalized eigenvalue problem to solve the Schr\"{o}dinger equation by the coupled-channel basis (\ref{basis3}).
For the spin singlet $|00 \rangle$ and longitudinal component $|10 \rangle$ of the spin triplet channels, the Hamiltonian (\ref{H_rel2}) can be decomposed into the diagonal and off-diagonal parts:
\begin{equation}
\hat{H}=\hat{H}_{\rm{diag}}+\hat{H}_{\rm{off}},
\end{equation}
where $\hat{H}_{\rm{off}} = -(\bm{\mu}_1+\bm{\mu}_2) \cdot \bm{B} $ and $\braket{00|\hat{H}_{\rm{off}}|10} = \braket{10|\hat{H}_{\rm{off}}|00}\not=0$.
The Schr\"{o}dinger equation for the coupled channel is written as
\begin{equation}
(\hat{H}_{\rm{diag}} + \hat{H}_{\rm{off}}) \Psi (\rho,z,\phi) = E \Psi (\rho,z,\phi).
\end{equation}
In the CGEM, we choose Eq.~(\ref{basis3}) as a trial wave function and then the Rayleigh-Ritz variational method leads to the generalized eigenvalue problem as follows:
\begin{widetext}
\begin{eqnarray}
&& \left(
\begin{array}{cc}
\left(
\begin{array}{ccc}
\bra{ \Phi^{P}_1}\hat{H}_{\rm{diag}}\ket{ \Phi^{P}_1} & \cdots & \bra{ \Phi^{P}_1}\hat{H}_{\rm{diag}}\ket{ \Phi^{P}_N} \\
\vdots & \ddots &  \\
\bra{ \Phi^{P}_N}\hat{H}_{\rm{diag}}\ket{ \Phi^{P}_1} &        & \bra{ \Phi^{P}_N}\hat{H}_{\rm{diag}}\ket{ \Phi^{P}_N} \\
\end{array}
\right)
&
\left(
\begin{array}{ccc}
\bra{ \Phi^{P}_{1}}\hat{H}_{\rm{off}}\ket{ \Phi^{V}_{N+1}} & \cdots & \bra{ \Phi^{P}_{1}}\hat{H}_{\rm{off}}\ket{ \Phi^{V}_{2N}} \\
\vdots & \ddots &        \\
\bra{ \Phi^{P}_{N}}\hat{H}_{\rm{off}}\ket{ \Phi^{V}_{N+1}} &        & \bra{ \Phi^{P}_N}\hat{H}_{\rm{off}}\ket{ \Phi^{V}_{2N}}
\end{array}
\right) \\
\left(
\begin{array}{ccc}
\bra{ \Phi^{V}_{N+1}}\hat{H}_{\rm{off}}\ket{ \Phi^{P}_{1}} & \cdots & \bra{ \Phi^{V}_{N+1}}\hat{H}_{\rm{off}}\ket{ \Phi^{P}_{N}}\\
\vdots & \ddots &         \\
\bra{ \Phi^{V}_{2N}}\hat{H}_{\rm{off}}\ket{ \Phi^{P}_{1}} &        & \bra{ \Phi^{V}_{2N}}\hat{H}_{\rm{off}}\ket{ \Phi^{P}_{N}}
\end{array}
\right)
&
\left(
\begin{array}{ccc}
\bra{ \Phi^{V}_{N+1}}\hat{H}_{\rm{diag}}\ket{ \Phi^{V}_{N+1}} & \cdots & \bra{ \Phi^{V}_{N+1}}\hat{H}_{\rm{diag}}\ket{ \Phi^{V}_{2N}} \\
\vdots & \ddots & \\
\bra{ \Phi^{V}_{2N}}\hat{H}_{\rm{diag}}\ket{ \Phi^{V}_{N+1}} &        & \bra{ \Phi^{V}_{2N}}\hat{H}_{\rm{diag}}\ket{ \Phi^{V}_{2N}}\\
\end{array}
\right) \\
\end{array}
\right)
\left(
\begin{array}{c}
C^{P}_1 \\
\vdots \\
C^{P}_N \\
\\
C^{V}_{N+1} \\
\vdots \\
C^{V}_{2N} \\
\end{array}
\right) \nonumber \\
&& \hspace{45pt} = E 
\left(
\begin{array}{cc}
\left(
\begin{array}{ccc}
\braket{ \Phi^{P}_1 | \Phi^{P}_1} & \cdots & \braket{ \Phi^{P}_1 | \Phi^{P}_N} \\
\vdots & \ddots &  \\
\braket{ \Phi^{P}_N | \Phi^{P}_1} &        & \braket{ \Phi^{P}_N | \Phi^{P}_N} \\
\end{array}
\right)
& 0 \\
0 & 
\left(
\begin{array}{ccc}
\braket{ \Phi^{V}_{N+1} | \Phi^{V}_{N+1}} & \cdots & \braket{ \Phi^{V}_{N+1} | \Phi^{V}_{2N}} \\
\vdots & \ddots & \\
\braket{ \Phi^{V}_{2N} |\Phi^{V}_{N+1}} &         & \braket{ \Phi^{V}_{2N} | \Phi^{V}_{2N}} \\
\end{array}
\right) \\
\end{array}
\right)
\left(
\begin{array}{c}
C^{P}_1 \\
\vdots \\
C^{P}_N \\
\\
C^{V}_{N+1} \\
\vdots \\
C^{V}_{2N} \\
\end{array}
\right),
\end{eqnarray}
\end{widetext}
where the superscripts $P$ and $V$ stand for $S=0$ and $S=1$, respectively.
Components $C_1, C_2 \cdots$ of the eigenvectors are coefficients of the basis of a trial wave function in Eq.~(\ref{basis3}).

\section{Charmonium excited states from effective Lagrangian} \label{App_EFT}
Here we briefly follow the approach based on the effective Lagrangian suggested in Refs.~\cite{Cho:2014exa,Cho:2014loa} and discuss the excited states of charmonia in a magnetic field from the point of view of the hadron degrees of freedom (see also Ref.~\cite{Gubler:2015qok} for $D$ mesons).
In this approach, the effective interaction Lagrangian with vector and pseudoscalar fields and the dual field strength tensor is given by 
\begin{equation}
\mathcal{L}_{\gamma \mathrm{PV}} = \frac{g_\mathrm{PV}}{m_0} e\tilde{F}_{\mu\nu} (\partial^\mu P) V^\nu,
\end{equation}
where $m_0=(m_P+m_V)/2$ is the averaging mass with $m_P$ and $m_V$ in vacuum, and $g_\mathrm{PV}$ is the dimensionless coupling constant.

Assuming a magnetic field along the $z$-direction, $\tilde{F}_{03} = - \tilde{F}_{30} = B$, and a zero spatial momentum $q_\mu =(\omega,0,0,0)$, the equations of motion with the mixing between pseudoscalar and longitudinal vector fields can be written as a matrix form. 
In particular, focusing on the mixing between $\eta_c$ and longitudinal $J/\psi$, the corresponding equations are written as the $2 \times 2$ ``partial'' matrix form \cite{Cho:2014exa,Cho:2014loa}:
\begin{equation}
\left(
\begin{array}{cc}
- \omega^{2} +  m_{\eta_c}^{2}  & - i \frac{g_1}{m_1} \omega eB \\
i \frac{g_1}{m_1} \omega eB & - \omega^{2} +  m_{J/\psi}^2 \\
\end{array}
\right)
\left(
\begin{array}{c}
\eta_c \\
{J/\psi}_{| 10 \rangle} \\
\end{array}
\right)
=0, \label{HadronEFT1}
\end{equation}
where we replaced $g_\mathrm{PV} \to g_1$ and $m_0 \to m_1 = (m_{\eta_c} + m_{J/\psi})/2$.

Similarly, for the mixing between $\eta_c(2S)$ and longitudinal $\psi(2S)$, we can find
\begin{equation}
\left(
\begin{array}{cc}
- \omega^{2} +  m_{\eta_c(2S)}^{2}                 & - i \frac{g_4}{m_4} \omega eB \\
i \frac{g_4}{m_4} \omega eB & - \omega^{2} +  m_{\psi(2S)}^{2} \\
\end{array}
\right)
\left(
\begin{array}{c}
\eta_c(2S) \\
{\psi(2S)}_{| 10 \rangle} \\
\end{array}
\right)
=0, \label{HadronEFT2}
\end{equation}
where $m_4 = (m_{\eta_c(2S)} + m_{\psi(2S)})/2$.
The solution of the $2 \times 2$ matrix form is given in Refs.~\cite{Cho:2014exa,Cho:2014loa}.

Furthermore, including also the effective vertex with different radial excitations such as $\eta_c$-$\psi(2S)$ and $\eta_c(2S)$-$J/\psi$, we finally obtain the $4 \times4$ ``full" matrix form (below the $D \bar{D}$ threshold):
\begin{widetext}
\begin{equation}
\left(
\begin{array}{cccc}
- \omega^{2} +  m_{\eta_c}^{2} & - i \frac{g_1}{m_1} \omega eB & 0 & - i \frac{g_2}{m_2} \omega eB\\
i \frac{g_1}{m_1} \omega eB & - \omega^{2} +  m_{J/\psi}^{2} & - i \frac{g_3}{m_3} \omega eB & 0 \\
0 & i \frac{g_3}{m_3} \omega eB & - \omega^{2} +  m_{\eta_c(2S)}^{2} & - i \frac{g_4}{m_4} \omega eB \\
i \frac{g_2}{m_2} \omega eB & 0 & i \frac{g_4}{m_4} \omega eB & - \omega^{2} +  m_{\psi(2S)}^{2}  \\
\end{array}
\right)
\left(
\begin{array}{c}
\eta_c \\
{J/\psi}_{| 10 \rangle} \\
\eta_c(2S) \\
{\psi(2S)}_{| 10 \rangle} \\
\end{array}
\right)
=0, \label{HadronEFT3}
\end{equation}
\end{widetext}
where $m_2 = (m_{\eta_c} + m_{\psi(2S)})/2$ and $m_3 = (m_{\eta_c(2S)} + m_{J/\psi})/2$.
Then we emphasize that this form partially includes not only Eqs.~(\ref{HadronEFT1}) and (\ref{HadronEFT2}) but also the new mixing terms being proportional to $g_2$ and $g_3$.

Next, let us determine the coupling constants, $g_1$, $g_2$, $g_3$ and $g_4$.
According to Appendix A in Ref.~\cite{Cho:2014loa}, the coupling constants can be determined by the experimental values of the radiative decay widths of charmonia:
\begin{equation}
g_\mathrm{PV} = \sqrt{12 \pi e^{-2} \tilde{p}^{-3} m_0^2 \Gamma_\mathrm{rad}}, \label{gpv}
\end{equation}
where $e=\sqrt{4\pi \alpha_{\mathrm{em}}}$ and $\tilde{p} = (m_i^2 - m_f^2)/(2m_i)$ is the center-of-mass momentum in the final state.
Then, we input the experimental values of $\Gamma_{J/\psi \to \gamma \eta_c} =1.579 \, \mathrm{keV}$, $\Gamma_{\psi(2S) \to \gamma \eta_c} = 1.0166 \, \mathrm{keV} $ and $\Gamma_{\psi(2S) \to \gamma \eta_c(2S)} = 0.2093\, \mathrm{keV}$ \cite{Agashe:2014kda}, and $\Gamma_{\eta_c(2S) \to \gamma J/\psi} = 15.7\, \mathrm{keV}$ from the lattice QCD simulations \cite{Becirevic:2014rda}.
Note that the radiative decay $\eta_c(2S) \to \gamma J/\psi$ is still not measured and the predicted values are controversial (e.g. $1.53 \, \mathrm{keV}$ \cite{Ebert:2002pp}, $5.6$-$7.9 \, \mathrm{keV} $ \cite{Barnes:2005pb} and $1.64 \, \mathrm{keV}$ \cite{Deng:2015bva}).
It is worthwhile to comment that the uncertainty of $\Gamma_{\eta_c(2S) \to \gamma J/\psi}$ (and $g_3$) can quantitatively affect the strength of the avoided crossing between the second and third states.
After substituting these values into Eq.~(\ref{gpv}), we arrive at the coupling strengths
\begin{eqnarray}
&& g_1 = g_{\gamma \eta_c J/\psi} =  2.0877,\\
&& g_2 = g_{\gamma \eta_c \psi(2S)} = 0.1346, \\
&& g_3 = g_{\gamma \eta_c(2S) J/\psi} = 0.7610, \\
&& g_4 = g_{\gamma \eta_c(2S) \psi(2S)} = 3.3762.
\end{eqnarray}

\begin{figure}[t!]
    \begin{minipage}[h]{1.0\columnwidth}
        \centering
        \includegraphics[clip, width=1.0\columnwidth]{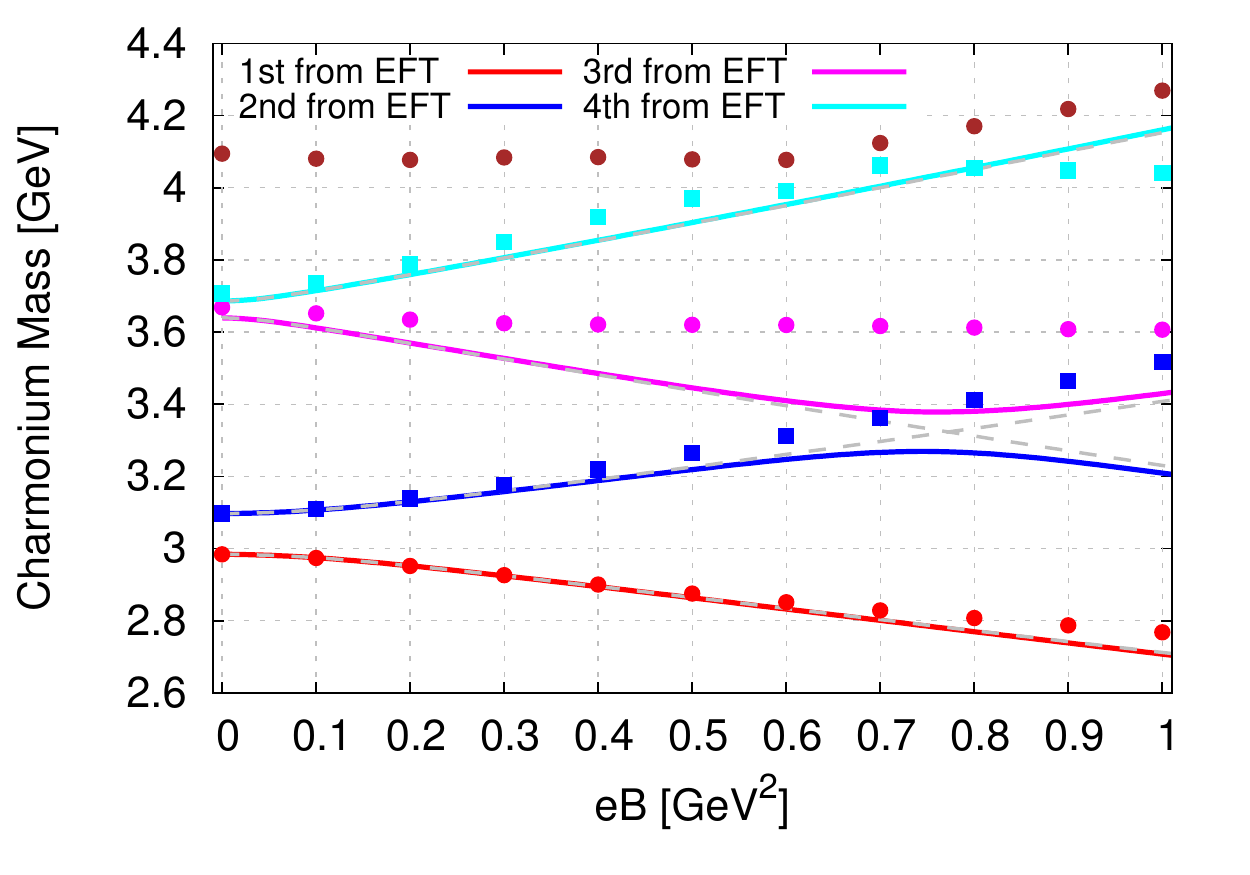}
    \end{minipage}
    \caption{Masses of $\eta_c (1S,2S)$ and longitudinal $J/\psi$, $\psi(2S)$ in a magnetic field {\it from the hadron effective Lagrangian}. 
The dashed lines are obtained from the partial matrix forms in Eqs.~(\ref{HadronEFT1}) and (\ref{HadronEFT2}).
The solid lines are obtained from the full matrix form in Eq.~(\ref{HadronEFT3}).
The dots are our results evaluated by the constituent quark model, Eq.~(\ref{H_rel2}), and CGEM.}
    \label{HadronEFT}
\end{figure}

In Fig.~\ref{HadronEFT}, we plot the mass formulas obtained from Eqs.~(\ref{HadronEFT1}), (\ref{HadronEFT2}), and (\ref{HadronEFT3}).
Here the mass shifts from the effective Lagrangian tend to be smaller than those from the quark model.
In particular, the crossing magnetic field between the second and third states is located in $eB=0.8 \, \mathrm{GeV}^2$, which is somewhat smaller than $eB=1.0$-$1.1 \, \mathrm{GeV}^2$ evaluated in the quark model.
One of the reasons is the absence of the effects based on quark degrees of freedom in the effective Lagrangian approach.
Then charmonia are considered as a point particle and the spacial information such as the $B$-dependence of the wave function is lost.
As a result, the predicted masses could be underestimated.
In addition, although the masses of the fourth state seem to be consistent in Fig.~\ref{HadronEFT}, the result from the effective Lagrangian dose not take into account the mixing with the fifth [or $\eta_c(3S)$] state, so that actually it would be overestimation because of the absence of the level repulsion and crossing.

\section{Magnetic field dependence of constituent quark mass from NJL model} \label{App_NJL}
In this appendix we mention briefly the magnetic field dependence of constituent quark masses.
One sees more details in Ref.~\cite{Boomsma:2009yk}.
Here, we do not take into account the effect of instantons.
To obtain the mean-field thermodynamical potential, we linearize the two-flavor NJL Lagrangian:
\begin{eqnarray}
\mathcal{L}_{\rm{NJL}} &=& \bar{\psi}(i\gamma^{\mu}\partial_{\mu}-m) \psi + G[( (\bar{\psi}\tau_a\psi)^2+ (\bar{\psi}\tau_ai\gamma_5\psi)^2)]\nonumber\\
&\simeq& \bar{\psi}(i\gamma^{\mu}\partial_{\mu}-\mathcal{M}) \psi+\frac{(M_0-m)^2}{4G}-\frac{M_3^2}{4G},
\end{eqnarray}
where $m=m_u=m_d$, $G$ and $\tau_a$ are the current light quark mass, coupling constant and Pauli matrix, respectively.
$\mathcal{M} = M_0 \tau_0 + M_3 \tau_3$, $M_0 = m-2G\braket{\bar{\psi}\tau_0\psi}$ and $M_3=m-2G\braket{\bar{\psi}\tau_3\psi}$.
Here, we assume the $\braket{\bar{\psi}\tau_0\psi}$ and $\braket{\bar{\psi}\tau_3\psi}$ condensates are nonzero.
The quadratic terms of the fluctuations are neglected.
In this approximation the thermodynamic potential is given by
\begin{eqnarray}
&& \Omega_{\rm{tot}}=\frac{(M_0-m)^2}{4G}-\frac{M_3^2}{4G} \\ \label{Thermo}
&&  - TN_c\sum_{f=u,d} \sum_{p_0=(2n+1) \pi T} \int\frac{d^3p}{(2\pi)^3} \ln \; \det [i \gamma_0 p_0 + \gamma_i p_i - M_f ], \nonumber
\end{eqnarray}
where $T$ is the temperature of the system and $N_c$ is the number of colors.
$M_u =M_0 + M_3$ and $M_d =M_0 - M_3$ are the constituent quark masses.
For the system in a finite magnetic field $B$, the dispersion relation of a quark and the integral over the three-momentum are modified as follows:
\begin{eqnarray}
&&p^2_{0,n}=p^2_z+M_f^2+(2n+1-s)|q_f|B\\
&&\int\frac{d^3p}{(2\pi)^3} \rightarrow\frac{|q_f|B}{2\pi} \sum_{n=0}^\infty\int\frac{dp_z}{2\pi},
\end{eqnarray}
where $n$ , $s$ and $q_f$ are the quantum number in the Landau levels, the spin and charge of the quark, respectively.

\begin{figure}[t!]
    \begin{minipage}[h]{1.0\columnwidth}
        \centering
        \includegraphics[clip, width=1.0\columnwidth]{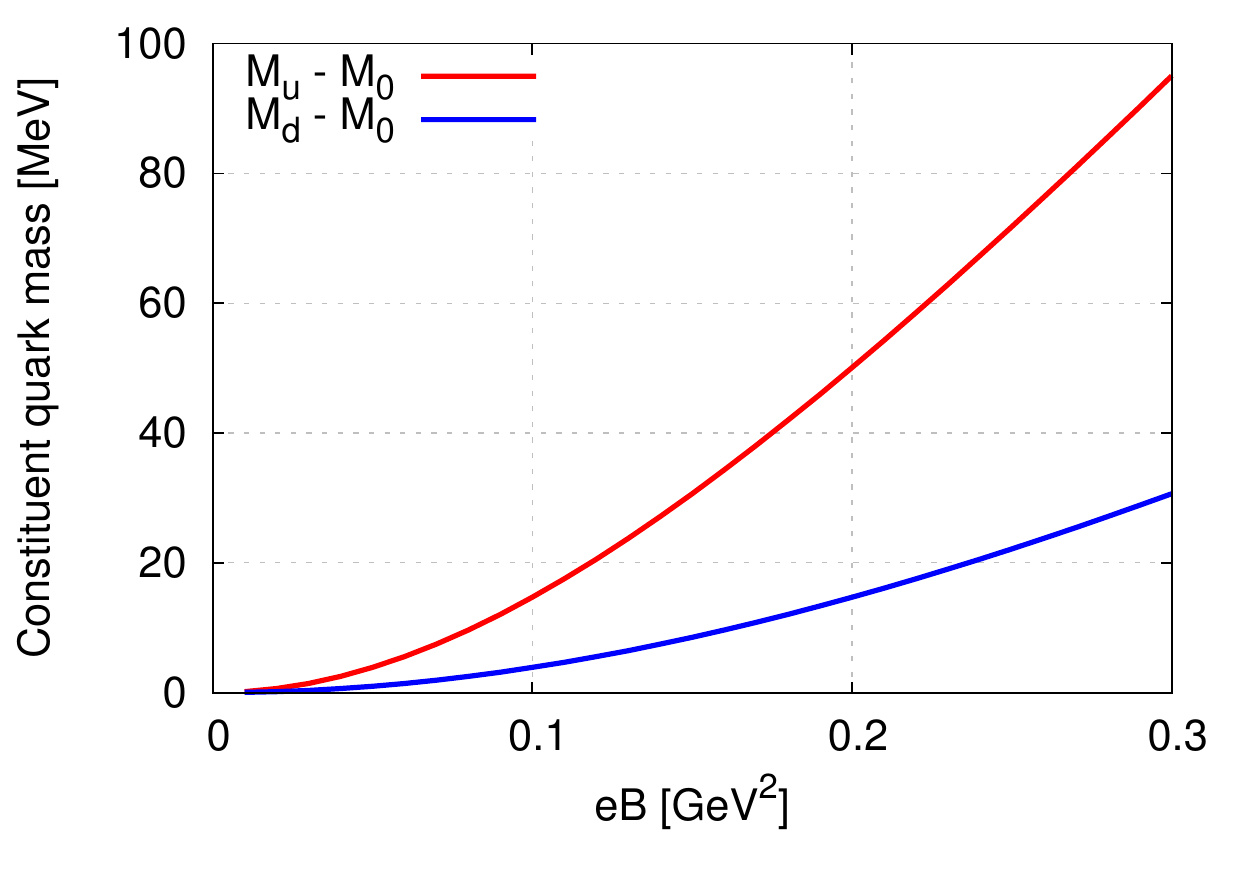}
    \end{minipage}
    \caption{Magnetic field dependences of the masses, $M_u$ and $M_d$, of light constituent quarks.
$M_0$ is the constituent quark mass in vacuum.}
    \label{MF_Dep_u_d}
\end{figure}

Then, the thermodynamic potential takes the form in the case of $T=0, B\not=0$:
\begin{widetext}
\begin{eqnarray}
\Omega_{\rm{tot}} &=& \Omega_0+\Omega_1 \nonumber\\
&=& \frac{(M_0-m)^2}{4G} - \frac{M_3^2}{4G}- \frac{N_c}{8\pi^2}\sum_{f=u,d} |M_f| \left[ M_f^3 \ln \left[ \frac{\Lambda}{M_f}+\sqrt{1+\frac{\Lambda^2}{M_f^2}} \right] - \Lambda(M_f^2 + 2 \Lambda^2) \sqrt{\frac{\Lambda^2}{M_f^2} + 1} \right] \nonumber \\
 &-& \frac{N_c}{2\pi^2} \sum_{f=u,d} (|q_f|B)^2[\zeta^{\prime}(-1,x_f) - \frac{1}{2}(x_f^2-x_f)\ln x_f+\frac{x_f^2}{4}],
\end{eqnarray}
\end{widetext}
where $x_f=\frac{M_f^2}{2|q_f|B}$ and $\zeta^{\prime}(-1,x_f)=\frac{d\zeta(z,x_f)}{dz}|_{z=-1}$ with the Hurwitz zeta function, $\zeta(z,x_f)$, and a conventional three-dimensional UV cutoff scheme is used.  $M_u$ and $M_d$ determined by minimizing the thermodynamical potential correspond to the masses of constituent quarks.
Magnetic field dependences on the constituent quark masses are shown in Fig.~\ref{MF_Dep_u_d} where we use $\Lambda=590 \, \rm{MeV}$ and $G\Lambda^2=2.435$ which are the same values as Ref.~\cite{Boomsma:2009yk}.

\bibliography{specto}
\end{document}